\documentclass{emulateapj}
\usepackage{amsmath}
\usepackage{amssymb}
\usepackage{natbib}
\usepackage{graphicx}
\usepackage{color}
\usepackage{hyperref}

\shorttitle{The Interplay between Spin, Mass, \& Morphology}
\shortauthors{A. F. Teklu et al.}

\begin{document}

\title{Connecting Angular Momentum and Galactic Dynamics:
The Complex Interplay between Spin, Mass, and Morphology}

\author{Adelheid F. Teklu$^{1,2}$, Rhea-Silvia Remus$^1$, Klaus Dolag$^{1,3}$, Alexander M. Beck$^1$, \\ Andreas Burkert$^{1,4}$, Andreas S. Schmidt$^3$, Felix Schulze$^1$, \& Lisa K. Steinborn$^1$}
\affil{$^1$ Universit\"ats-Sternwarte M\"unchen, Scheinerstra{\ss}e 1, D-81679 M\"unchen, Germany\\
$^2$ Excellence Cluster Universe, Boltzmannstra{\ss}e 2, D-85748 Garching, Germany\\
$^3$ Max-Planck Institute for Astrophysics, Karl-Schwarzschild-Str. 1, D-85741 Garching, Germany\\
$^4$ Max-Planck Institute for Extraterrestrial Physics, Giessenbachstra{\ss}e 1, D-85748 Garching, Germany \\ \texttt{ateklu@usm.lmu.de} \\}


\begin{abstract}
 The evolution and distribution of the angular momentum of dark matter (DM)
 halos have been discussed in several studies over the past decades. In particular, the idea arose that angular momentum
 conservation should allow to infer the total angular momentum of the entire DM halo from measuring the angular
 momentum of the baryonic component, which is populating the center of
 the halo, especially for disk galaxies. To test this idea and to
 understand the connection between the angular momentum of the DM halo and its galaxy, we use a state-of-the-art, hydrodynamical
 cosmological simulation taken from the set of Magneticum Pathfinder simulations.  Thanks to the inclusion of the relevant physical
 processes, the improved underlying numerical methods, and high spatial
 resolution, we successfully produce populations of
   spheroidal and disk galaxies self-consistently. Thus, we are able
   to study the dependence of galactic properties on their
   morphology.  We find that (1) the specific angular momentum of stars in disk and spheroidal galaxies 
as a function of their stellar mass compares well with observational 
results;
(2) the specific angular
 momentum of the stars in disk galaxies is slightly smaller
 compared to the specific angular momentum of the cold gas, in good
 agreement with observations;
(3) simulations including the baryonic
 component show a dichotomy in the specific stellar angular momentum
 distribution when splitting the galaxies according to their
 morphological type (this dichotomy can also be seen in the spin
 parameter, where disk galaxies populate halos with slightly larger
 spin compared to spheroidal galaxies); (4) disk galaxies
 preferentially populate halos in which the angular momentum vector of
 the DM component in the central part shows a better
 alignment to the angular momentum vector of the
 entire halo; and (5) the specific angular momentum of
 the cold gas in disk galaxies is approximately 40\%
 smaller than the specific angular momentum of the total DM halo and shows a significant scatter.  

\end{abstract}



\section{Introduction}

The question of how the different types of galaxies formed, and, in particular, which properties determine a galaxy to become a disk or spheroidal, has been a matter of debate since the discovery that the Universe is full of distant stellar islands of different morphology. 
For elliptical galaxies the following formation scenarios are discussed: 
from observations of ongoing mergers between galaxies of similar masses we know that spheroids can form in a major merger event between (spiral) galaxies. This has been supported by many simulations (e.g., \citealt{Toomre77,white78,white1979,barnes1996,naab2006}). However, this scenario is not sufficient to explain certain observational properties, especially for the most massive galaxies. Thus, an alternative formation scenario through a series of multiple minor merger events was proposed by \citet{Meza03} and established by, e.g., \citet{naab2009}, \citet{gonzales2009}, \citet{oser2010}, and \citet{johansson2012}.

For the formation scenarios of spiral galaxies, the details of their formation processes are less well known, but all the different channels discussed in the literature over the past decades are connected to the detailed buildup of the angular momentum and how the gaseous component transports intrinsic angular momentum
into the central part of the halo, where the galaxy assembles.
Initially, the dark matter (DM) halo and the infalling gas have identical angular momenta, but during the formation of the galaxy the gas cools and condenses in the center of the halo to form a disk.
Assuming that the angular momentum of the gas is conserved during this process, the angular momentum of the halo of a disk galaxy can be estimated directly from the angular momentum of the disk of the galaxy \citep{FE80,Fall83,MMW98}.

In hierarchical scenarios of structure formation, structures form through the gathering of clumps owing to the gravitational force 
\citep[][and references therein]{Peebles93}. 
The DM collapses at high redshifts into small objects, which grow into larger objects through merging and finally build halos. 
Those halos are not completely spherically symmetric owing to tidal torques induced by neighboring halos.
The baryons condense in the centers of these DM halos and form the first protogalaxies \citep{WR78}. Since the baryons and the DM originally gained the same amount of angular momentum through these tidal torques \citep{Peebles69,Doro70} and the angular momentum of the gas should be conserved during the collapse, the disk is expected to have a similar specific angular momentum to that of its hosting halo \citep{FE80}. 
 
In contrast to observations, numerical $N$-body simulations have the advantage that structures can be followed through time, enabling the detailed understanding of the early stages of galaxy formation. 
Until recently, simulations using traditional smoothed particle hydrodynamics (SPH) codes suffered from the so-called angular momentum problem, where the objects became too small compared to observations because the gas had lost too much angular momentum (e.g., \citealp{NB91,NW94,NS97}). 
\citet{Sijacki12} showed that SPH simulations overestimate the number of elliptical galaxies owing to the lower amount of mixing, and therefore causing this spurious ``angular momentum crisis'' of the baryons. On the opposite side, adaptive mesh refinement (AMR) or moving mesh simulations tend to overpredict the amount of disk-like galaxies in the absence of any feedback \citep{Scannapieco12}. 
As simulations become better in resolution and also include feedback from stars, supernovae (SNe), and active galactic nuclei (AGNs), the gas is prevented from cooling too soon. 
Hence, early star formation is suppressed and the associated loss of angular momentum can be minimized. 
There are several studies that investigate the influence of star formation and the associated SN feedback on the formation of galactic disks (e.g., \citealp{Brook04,Okamoto05,Governato07,Scan08,Zavala08}). 
They find that strong feedback at early times leads to the formation of more realistic disk galaxies.
A more recent study on the effect of stellar feedback on the angular momentum is presented in \citet{Uebler14}, where they found that strong feedback favors disk formation. A recent, detailed summary of disk galaxy simulations can be found in \citet{Murante2015}. 

One parameter to describe the rotation of a system is the so-called spin parameter $\lambda$, which was introduced by \citet{Peebles69,Peebles71} and has since been investigated in several studies. 
With $\lambda$ it is possible to measure the degree of rotational support of the total halo. 
It has the advantage that it is only weakly depending on the halo's mass and its internal substructures \citep{BE87}. 
The connection of this parameter with galaxy formation has been studied by many authors.
 
At first, DM-only simulations were employed. \citet{B01} introduced a modified version of the $\lambda$-parameter by defining the energy of the halo via its circular velocity. In addition, they studied the alignment of the angular momentum of the inner and outer halo parts. They found that most halos were well aligned but that about 10\% of the halos showed a misalignment. 
\citet{Aubert04} investigated the alignment between the inner spin of halos and the angular momentum of the outer halo and also the alignment between the inner spin and the angular momentum of the inflowing material. 
The misalignment angle of the angular momentum vectors at different radii was studied in more detail by \citet{BS05}, who showed that with increasing separation of the radii the misalignment of the vectors increases. In another study, \citet{Maccio08} focused on the spin parameter's dependency on the mass and the cosmology and found no correlations. \citet{Trowland13} studied the connection of the halo spin with the large-scale structure. They confirmed that the spin parameter is not mass dependent at low redshift but found a tendency to smaller spins at higher redshifts.

Other authors included nonradiative gas in their analysis. The misalignment of the gas and DM angular momentum vectors and their spin parameters at redshift $z=3$ were investigated by \citet{vdB02}. They found the angular momentum vectors to be misaligned by a median angle of about $30^{\circ}$. However, the overall distributions of the spin parameters of the gas and the DM were found to be very similar. 
Another detailed study by \citet{SS05} and \citet{SSBH12} found that the spin parameter of the gas component is on average higher than that of the DM, and they reported a misalignment of the angular momentum of the gas and the DM of about $20^{\circ}$. 
 
\citet{Chen03} compared the spin parameter and the misalignment angles between the DM and gas angular momentum vectors obtained in simulations, which include radiative cooling.
This enables a splitting of the gas into a cold and a hot component.
In their nonradiative model they confirmed that the spin parameter of the gas component has higher spin than the DM, while in their simulations with cooling the two components had approximately the same spin. 
The misalignments of the global angular momentum vectors were $22.8^{\circ}$ and $25^{\circ}$ for the two different cooling models and $23.5^{\circ}$ for the nonradiative case.
\citet{Stewart11,Stewart13} focused on the specific angular momentum and the spin parameter in relation to the cold/hot mode accretion by following the evolution of individual halos over cosmic time. They found that the spin of cold gas was the highest compared to all other components and about $3-5$ times higher than that of the DM component.
\citet{Scan09} investigated the evolution of eight individual halos from zoom simulations and did not find a correlation between the spin parameter and the morphology of the galaxy. However, they found a correlation between the morphology and the specific angular momentum: for disks, the specific angular momentum can be higher than that of the host halo, while spheroids tend to have lower specific angular momentum. 
In addition, they saw a misalignment between stellar disks and infalling cold gas. 
Also, \citet{Sales12} reported no correlation between the spin parameter and the galaxy type and concluded that disks are predominantly formed in halos where the freshly accreted gas has similar angular momentum to that of earlier accretion, whereas spheroids tend to form in halos where gas streams in along misaligned cold flows. 
The angular momentum properties of the inflowing gas were studied by \citet{Pichon11}. They found that the angular momentum of the cold dense gas is well aligned with the angular momentum of the DM halo, in contrast to the hot diffuse gas. Additionally, an increase of the advected specific angular momentum of the gas component with cosmic time was noticed. They proposed a scenario in which the angular momentum of the galaxies is fed by the collimated cold flows that are amplified with time and make the disks larger. 
\citet{Kimm11} studied the different behavior of the gas and DM specific angular momentum by following the evolution of a Milky-Way-type galaxy over cosmic time. 
They also found that the gas has higher specific angular momentum and spin parameter than the DM. 
\citet{Hahn10} investigated a sample of about 100 galactic disks and their alignment with their host halo at three different redshifts. 
Both the stellar and gas disks had a median misalignment angle of about $49^{\circ}$ with respect to the hosting DM halo at $z=0$. 
 
Some works focused explicitly on the effect that baryons have on the DM halo by running parallel DM-only simulations. 
\citet{Bryan13}, who extracted halos from OWLS \citep{Schaye10}, did not find a dependency of the spin parameter on mass, redshift, or cosmology in their DM-only runs. 
In the simulations that included baryons and strong feedback, the overall spin was found to be affected only very little, while the
baryons had a noticeable influence on the inner halo, independent of the feedback strength. 
\citet{Bett10} investigated the specific angular momentum and the misalignment of the galaxy with its halo. 
They obtained a median misalignment angle of about $25^{\circ}$ for the DM-only runs and about $30^{\circ}$ for the run including baryons. 
The baryons were found to spin up the inner region of the halo. 

Recently, \citet{Welker14} studied the alignment of the galaxy spins with their surrounding filaments, using the Horizon-AGN simulation \citep{Dubois14}. 
They find that halos experiencing major mergers often lower the spin, while in general minor mergers can increase the amount of angular momentum. 
If a halo does not undergo any mergers but only smooth accretion, the spin of the galaxy increases with time, in contrast to that of its hosting DM halo. 
Since the gas streams and clumps in general move along the filaments, the galaxies realign with their filaments. 
\citet{Danovich15} have investigated the buildup of the angular momentum in galaxies, using a sample of 29 resimulated galaxies at redshifts from $z=4$ to $1.5$. 
Overall the spin of the cold gas was about three times higher than that of the DM halo, in line with previous studies.
\citet{Genel15} explored the importance of stellar and AGN feedback for the evolution of the angular momentum of the galaxies in the Illustris simulations, finding the angular momentum of their simulated galaxies to be in good agreement with observations for a chosen feedback similar to the one used in this paper.

Observations indicate that the morphology of galaxies is strongly influenced by the relation between mass and angular momentum (see \citealp{Fall83}). 
The angular momentum of disk galaxies was found to be about six times higher than that of the ellipticals of equal mass. 
\citet{Romanowsky12} and \citet{Fall13} revisited and extended this work, analyzing 67 spiral and 40 early-type galaxies. 
They found that lenticular (S0) galaxies lie between spiral and elliptical galaxies in the so-called $M_\mathrm{star}-j_\mathrm{star}$-plane. 
The bulges of spiral galaxies follow a similar relation because they behave like ``mini-ellipticals.'' 
\citet{Obreschkow14} used data from high-precision measurements of 16 nearby spiral galaxies to calculate the specific angular momentum of the gas and the stars. 
They confirmed observationally that the mass and angular momentum are strongly correlated with the morphology of galaxies.
\citet{Hernandez06} calculated the $\lambda$-parameter for two galaxy samples with a total of 337 observed spirals and found that with decreasing bulge-to-disk ratio the spirals have increasing $\lambda$-values. In a further study \citet{Hernandez07} investigated a sample of 11,597 spiral and elliptical galaxies from the Sloan Digital Sky Survey (SDSS) and found that ellipticals on average have lower $\lambda$-values than spiral galaxies.

Recently, a new generation of cosmological, hydrodynamical simulations, e.g., the MassiveBlack-II \citep{Khandai15}, Magneticum Pathfinder \citep{Hirschm14}, Illustris \citep{Vogelsberger14}, Horizon-AGN \citep{Dubois14}, and EAGLE \citep{EAGLE15} simulations, have been employed to follow the evolution of structures in the Universe. 
A new aspect of such simulations is that for the first time reasonable galaxy morphologies can be associated with the galaxies formed in those simulations, 
e.g., Horizon-AGN \citep{Dubois14}, Illustris \citep{Torrey2015}, EAGLE \citep{EAGLE15}, and Magneticum Pathfinder \citep{Remus15}. In this work we will analyze simulations from the set of Magneticum Pathfinder simulations\footnote{www.magneticum.org} (K. Dolag et al., in preparation), which are introduced in Section 2, and investigate how the baryonic component influences the morphology of the galaxy. 
In Section 3 we introduce the formulae and investigate the angular momentum, the spin parameter, and the alignments of halos. 
In Section 4 we show the kinematical split-up of galaxies, which allows us to classify the galaxies in the simulation using the circularity parameter $\varepsilon$ in Section 5.
We then examine the angular momentum of the different components of spheroids and disks. 
Furthermore, we study the alignment of the angular momentum vectors of the baryonic and the DM component. 
Additionally, we analyze the differences in the spin parameter $\lambda$ of spheroidal and disk galaxies. 


\section{The Magneticum Pathfinder Simulations}\label{sec:sim}

In order to study the properties of galaxies in a statistically relevant manner, we need both a large sample size and high enough resolution to resolve the morphology and underlying physics of galaxies. 
Since the newest generation of cosmological simulations can achieve both, they are a valuable tool for this study. 
We take the galaxies for our studies from the Magneticum Pathfinder simulations, which are a set of cosmological hydrodynamical simulations with different volumes and resolutions (K. Dolag et al., in preparation). 
The simulations were performed with an extended version of the $N$-body/SPH code GADGET-3, which is an updated version of GADGET-2 \citep{Springel01a,Springel05gad}. 
It includes various updates in the formulation of SPH regarding the treatment of the viscosity and the used kernels (see \citealp{Dolag05,Donnert2013,Beck15}).

It also allows a treatment of radiative cooling, heating from a uniform time-dependent ultraviolet (UV) background, and star formation with the associated feedback processes. 
The latter is based on a subresolution model for the multiphase structure of the interstellar medium \citep{Springel03}. 
Radiative cooling rates are computed following the same procedure presented by \citet{Wiersma09}.
We account for the presence of the cosmic microwave background (CMB) and of UV/X-ray background radiation from quasars and
galaxies, as computed by \citet{Haardt01}. 
The contributions to cooling from each one of 11 elements (H, He, C, N, O, Ne, Mg, Si, S, Ca, Fe) have been precomputed using the publicly available CLOUDY photoionization code \citep{Ferland98} for an optically thin gas in (photo)ionization equilibrium. 

In the multiphase model for star formation \citep{Springel03}, the ISM is treated as a two-phase medium where clouds of cold gas form from cooling of hot gas and are embedded in the hot gas phase assuming pressure equilibrium whenever gas particles are above a given threshold density. 
The hot gas within the multiphase model is heated by SNe and can evaporate the cold clouds. 
A certain fraction of massive stars (10\%) is assumed to explode as SNe II. 
The released energy by SNe II ($10^{51}$~erg) is modeled to trigger galactic winds with a mass loading rate being proportional to the star formation rate (SFR) to obtain a resulting wind velocity of $v_{\mathrm{wind}} = 350$km/s.
Our simulations also include a detailed model of chemical evolution according to \citet{Tornatore07}. 
Metals are produced by SNe II, by SNe Ia, and by intermediate- and low-mass stars in the asymptotic giant
branch (AGB). 
Metals and energy are released by stars of different mass by properly accounting for mass-dependent lifetimes (with a lifetime function
according to \citealp{Padovani93}), the metallicity-dependent stellar yields by \citet{Woosley95} for SNe II, the yields by \citet{vandenHoek97} for AGB stars, and the yields by \citet{Thielemann03} for SNe Ia. 
Stars of different mass are initially distributed according to a Chabrier initial mass function \citep{Chabrier03}.

Most importantly, our simulations also include a prescription for black hole (BH) growth and for feedback from AGNs
based on the model presented in \citet{Springel05a} and \citet{DiMatteo05}, including the same modifications as in the study of \citet{Fabjan10} and some new, minor changes.

As for star formation, the accretion onto BHs and the associated feedback adopt a subresolution model. 
BHs are represented by collisionless ``sink particles'' that can grow in mass by accreting gas from their environments, or by merging with other BHs. 
The gas accretion rate $\dot{M}_\bullet$ is estimated using the Bondi-Hoyle-Lyttleton approximation (\citealp{Hoyle39, Bondi44,
  Bondi52}): 
\begin{equation}\label{Bondi}
\dot{M}_\bullet = \frac{4 \pi G^2 M_\bullet^2 f_{\mathrm boost} \rho}{(c_s^2 + v^2)^{3/2}},
\end{equation}
where $\rho$ and $c_s$ are the density and the sound speed of the surrounding (ISM) gas, respectively, $f_{\mathrm boost}$ is a boost factor for the density, which typically is set to $100$ and $v$ is the velocity of the BH relative to the surrounding gas. 
The BH accretion is always limited to the Eddington rate (maximum possible accretion for balance between inward-directed gravitational force and outward-directed radiation pressure): 
$\dot{M}_\bullet = \min(\dot{M}_\bullet, \dot{M}_{\mathrm{edd}})$. 
Note that the detailed accretion flows onto the BHs are unresolved, and thus we can only capture BH growth due to the larger-scale gas distribution, which is resolved. 
Once the accretion rate is computed for each BH particle, the mass continuously grows. 
To model the loss of this gas from the gas particles, a stochastic criterion is used to select the surrounding gas particles to be removed. 
Unlike in \citet{Springel05a}, in which a selected gas particle contributes with all its mass, we included the possibility for a gas particle to lose only a slice of its mass, which corresponds to 1/4 of its original mass. 
In this way, each gas particle can contribute with up to four `generations' of BH accretion events, thus providing a more continuous description of the accretion process.

The radiated luminosity $L_{\mathrm{r}}$ is related to the BH accretion rate by $L_{\mathrm{r}} = \epsilon_{\mathrm{r}} \dot{M}_\bullet c^2$, where $\epsilon_{\mathrm{r}}$ is the radiative efficiency, for which we adopt a fixed value of 0.1 (standardly assumed for a radiatively efficient accretion disk onto a nonrapidly spinning BH according to \citealp{Shakura73}, see also \citealp{Springel05gad,  DiMatteo05}). 
We assume that a fraction $\epsilon_{\mathrm{f}}$ of the radiated energy is thermally coupled to the surrounding gas so that $\dot{E}_{\mathrm{f}} = \epsilon_{\mathrm{r}} \epsilon_{\mathrm{f}} \dot{M}_\bullet c^2$ is the rate of the energy feedback; $\epsilon_{\mathrm{f}}$ is a free parameter and typically set to $0.1$ (see discussion in \citet{Steinborn15}). 
The energy is distributed kernel weighted to the surrounding gas particles in an SPH-like manner. 
Additionally, we incorporated the feedback prescription according to \citet{Fabjan10}: 
we account for a transition from a quasar- to a radio-mode feedback (see also \citealp{Sijacki07}) whenever the accretion rate falls below an Eddington ratio of $f_{\mathrm{edd}} := \dot{M}_{\mathrm{r}}/ \dot{M}_{\mathrm{edd}} < 10^{-2}$. 
During the radio-mode feedback we assume a 4 times larger feedback efficiency than in the quasar mode.
This way, we want to account for massive BHs, which are radiatively inefficient (having low accretion rates), but which are
efficient in heating the ICM by inflating hot bubbles in correspondence to the termination of AGN jets. 
Note that we also, in contrast to \citet{Springel05a}, modify the mass growth of the BH by taking into account the feedback, e.g.,
$ \Delta M_\bullet = (1-\eta_{r})\dot{M}_\bullet \Delta t$.
Additionally, we introduced some technical modifications of the original implementation, for which readers can find details in 
\citet{Hirschm14}, where we also demonstrate that the bulk properties of the AGN population within the simulation
are similar to observed AGN properties.

The simulation additionally follows thermal conduction, similar to \citet{Dolag04}, but with a choice of 1/20 of the classical Spitzer
value \citep{Spitzer62}.  
The choice of a suppression value significantly below 1/3 can be justified by comparison with full MHD simulations including an anisotropic treatment of thermal conduction (see discussion in \citet{Arth14}).

The initial conditions are using a standard $\Lambda$CDM cosmology with parameters according to the seven-year results of the \textit{Wilkinson Microwave Anisotropy Probe} (\textit{WMAP7}; \citealt{Komatsu11}). 
The Hubble parameter is $h=0.704$, and the density parameters for matter, dark energy, and baryons are $\Omega_{M}=0.272$, $\Omega_{\Lambda}=0.728$, and $\Omega_{b}=0.0451$, respectively. 
We use a normalization of the fluctuation amplitude at 8 Mpc of $\sigma_{8}=0.809$ and also include the effects of baryonic acoustic oscillations. 

The Magneticum Pathfinder simulations have already been successfully used in a wide range of numerical studies, showing good agreement with observational results for the pressure profiles of the intracluster medium \citep{Planck2013pp,SPT2014pp}, for the properties of the AGN population \citep{Hirschm14,Steinborn15}, and for the dynamical properties of massive spheroidal galaxies \citep{Remus13,Remus15}.

In this work we mainly used a medium-sized (48 Mpc/$h$)$^{3}$ cosmological box at the {\it uhr} resolution level, which initially contains a total of $2\cdot576^{3}$ particles (DM and gas) with masses of $m_\mathrm{DM} = 3.6\cdot10^{7} M_{\odot}/h$ and $m_\mathrm{gas} = 7.3\cdot10^{6} M_{\odot}/h$, having a gravitational softening length of 1.4 kpc/$h$ for DM and gas particles and 0.7 kpc/$h$ for star particles. Additionally, we performed a DM-only reference run, where we kept exactly the same initial conditions, e.g., the original gas particles were treated as collisionless DM particles.

To identify subhalos we used a version of SUBFIND \citep{Springel01b}, adapted to treat the baryonic component \citep{Dolag09}. 
SUBFIND detects halos based on a standard Friends-of-Friends algorithm \citep{Davis85} and self-bound subhalos around local density peaks within the main halos. 
The virial radius of halos is evaluated according to the density contrast based on the top-hat model \citep{Eke96}. 
In further post-processing steps we then extract the particle data for all halos and compute additional properties for the different components and within different radii, using the full (thermo) dynamical state of the different particle species.
In this study we mainly present results at redshifts $z = 2$, $z = 1$, $z = 0.5$, and $z = 0.1$.


\section{Properties of Halos}\label{sec:class}

From our data set we extract all halos with a virial mass above $5\cdot
10^{11}M_{\odot}$.  At a redshift of $z=2$ there are 396 halos, at
$z=1$ there are 606 halos, at $z=0.5$ we find 629 halos, and at $z=0.1$
there are 622 halos.  The lower limit is chosen to obtain a sample of
halos that contain a significant amount of stellar mass and for which
the resolution is sufficient to resolve the inner stellar, and gas
structures. We transform the positions and velocities of all gas,
stellar and DM particles into the frame of the halo, where the 
position given by SUBFIND is the position of the minimum of the potential 
and the velocity of the host halo is the mass-weighted mean velocity of all
particles belonging to the halo.

\begin{figure*}
\begin{centering}
\includegraphics[width=0.8\textwidth]{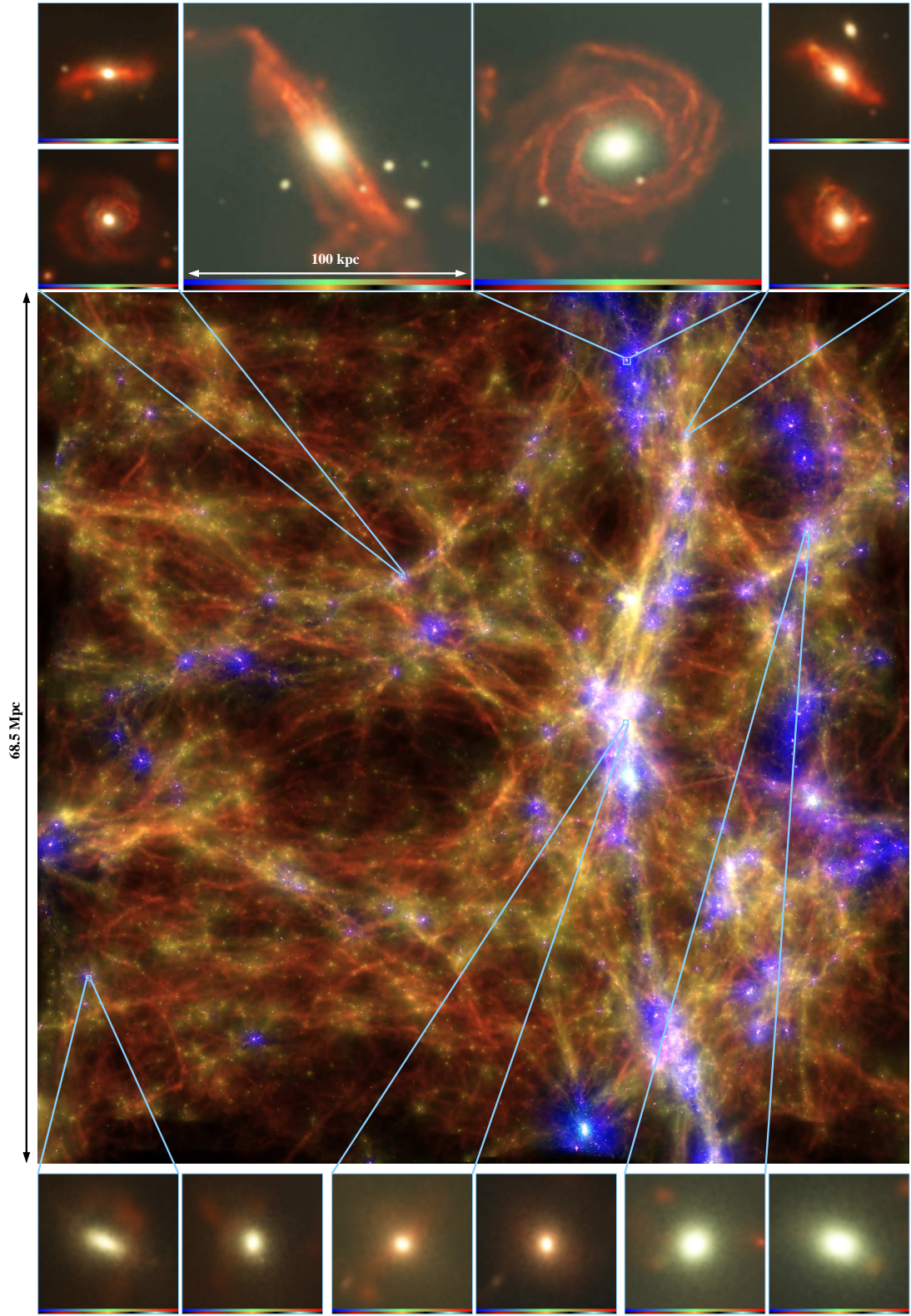}
\caption{Main panel: complete cosmological box at redshift $z = 0.5$. The upper panels show exemplary spiral
galaxies, and the lower panels show spheroidal galaxies.
We show two random projection directions for each of the galaxies.
In all panels we color-code the stellar particles by their cosmological formation epoch, where the upper color bars represent the age of the stars from old to young and the lower color bars represent the gas temperature from cold to hot. We also marked the position of the halos within the cosmological box.}
\label{fig:box4}
\end{centering}
\end{figure*}

Exemplary, Figure \ref{fig:box4} shows six galaxies within the chosen mass range demonstrating that the galaxies in the simulations look like observed disk and spheroidal galaxies. 
The middle panel is a picture of the full cosmological box at redshift $z=0.5$. 


\subsection{The Angular Momentum}\label{sec:angmom}

The mass and the angular momentum of galaxies are observed to be closely correlated with their morphology (e.g., \citealp{Fall83,Romanowsky12,Obreschkow14}). 
In a closed system without external forces, the angular momentum is a conserved quantity.
However, in the context of galaxy formation and the interaction between collapsing objects and the large-scale structure, the assumption of the conservation is not necessarily fulfilled for a single galaxy. 
The total angular momentum of a galactic halo is given by
\begin{equation} \textbf{J} = \sum_{k} \left(\sum_{i \in N_k} m_{k,i}\: \textbf{r}_{k,i} \times \textbf{v}_{k,i} \right),\end{equation}
where $k$ are the different particle types of our simulation (gas, stars, and DM) and $N_k$ is their corresponding particle number with the loop index $i$. 

In our simulations, each particle carries its own mass. The initial mass is different for gas, star, and DM particles. Later on,
the mass of individual gas particles varies owing to mass losses during
star formation. Thus, we remove the mass dependence and use the
specific angular momentum
\begin{equation} \label{eq:specangmom}
\textbf{j}_{k} = \frac {\sum_{i} m_{k,i} \textbf{r}_{k,i} \times
 \textbf{v}_{k,i}} {\sum_{i} m_{k,i}}, 
\end{equation}
where $k$ are the species of matter, as above. Therefore, we firstly
calculate the angular momentum of each particle of a species.
Afterward, we sum over all individual particles and divide by the
total mass of the corresponding species to obtain the absolute value. 


\subsection{The Spin Parameter $\lambda$}

In the following section we want to study the dimensionless
$\lambda$-parameter. 
As defined by \citet{Peebles69,Peebles71} and adopted by, e.g., \citet{MMW98}, the general $\lambda$-parameter, which is used for the total halo, is given by
\begin{equation} \label{eq:lambda}
\lambda = \frac{J |E|^{1/2}} {G M^{5/2}} ,
\end{equation}
where $E = - GM^2/2 \: R_\mathrm{vir}$ is the total energy of the
halo. 
This total spin parameter can only be used considering all matter inside the halo. To evaluate the different components (gas, stars, and DM), this parameter needs to be modified. 
We follow \citet{B01}, who defined a component-wise spin parameter as follows:
\begin{equation}
\lambda '(r) := \frac{J(r)}{\sqrt{2}M(r)V_\mathrm{circ}(r)r} ,
\end{equation}
where $J(r)$ is the angular momentum, $M(r)$ the mass, and
$V_\mathrm{circ} = \sqrt{GM(r)/r}$ the circular velocity within a
radius $r$. Another advantage of $\lambda '(r)$ is that it can be
used for the calculation at different radii. When calculated over the
entire virial\footnote{Formally, if the {\it true} energy $E$ is used, this is equal for a truncated, isothermal sphere. However, for a NFW halo the correction term is of order unity, see \citet{B01} and references therein} radius, $\lambda '(r)\approx \lambda$. 
For simplicity we will drop the prime for the remaining part of our study. 
For the evaluation of the different components $\lambda$ can be expressed in terms of the specific angular momentum, as done by \citet{vdB02}: 
\begin{equation}
\lambda_{k} = \frac{j_{k}}{\sqrt{2} R_\mathrm{vir} V_\mathrm{circ}} ,
\end{equation}
where $k$ stands for the different components. 

Independent of the definition, the distribution of the $\lambda$-parameter can be fitted by a
lognormal distribution of the following form:
\begin{equation} \label{eq:lambda_fit}
P(\lambda)=\frac{1}{\lambda \sqrt{2\pi} \sigma} \: \mathrm{exp}{\left(-
 \frac{ln^2(\lambda / \lambda_{0})}{2 \sigma^2}\right)},
\end{equation}
with the fit parameter $\lambda_{0}$, which is about the median
value, and the standard deviation $\sigma$.

\begin{figure}
\begin{centering}
\includegraphics[trim={0cm -0.5cm 0cm 0.5cm}, width=0.5\textwidth,clip=true]{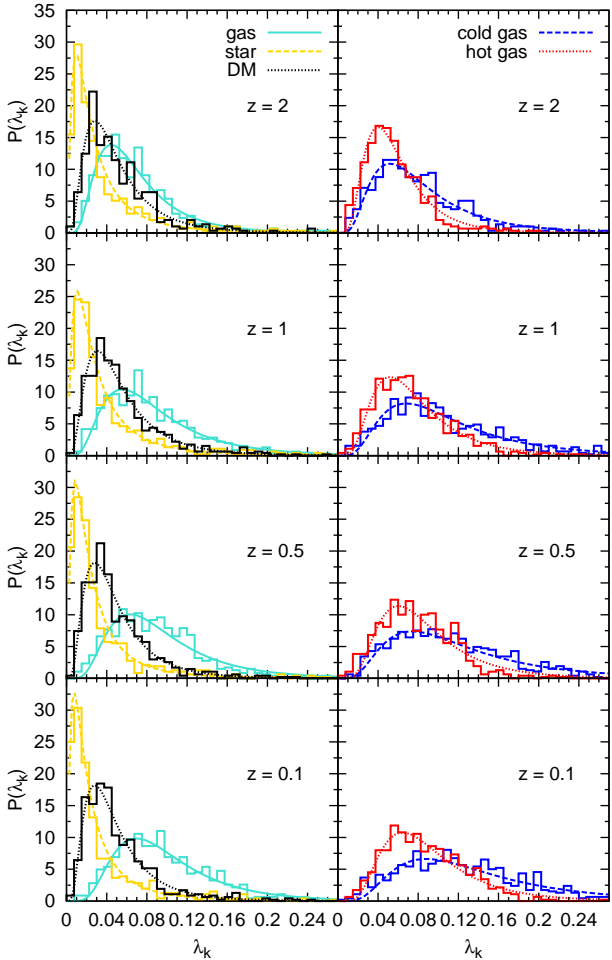}
\caption{Left panels: histograms for the $\lambda$-parameter for the
 different components within $R_\mathrm{vir}$; the DM (black) is the dominant component, the stars (yellow) peak at lower values, whereas the gas (turquoise) is distributed around higher values. While $\lambda_{0}$ stays relatively constant for the
 DM and stellar component, it increases for the gas with
 decreasing redshift. Right panels: the gas component splits into hot
 (red) and cold (blue) gas. Both components spin up with cosmic time,
 whereas for the cold gas this happens faster. Overplotted are the
 best-fit lognormal distributions (smooth lines), for which the
 parameters are reported in Table \ref{tab:lambda_comp}.}
\label{fig:lambda_comp}
\end{centering}
\end{figure}

The left panels of Figure \ref{fig:lambda_comp} show the histograms of the
$\lambda_k$-distribution in linear bins of all halos in the selected
mass range for stars (yellow), DM (black), and
gas (turquoise) from redshift $z=2$ (upper panels) to $z=0.1$
(lower panels). The histograms and fit curves (smooth lines) are each normalized to
the number of halos. 
The distribution of spin parameters of the stellar component has significantly lower values
compared to the distribution of the spin parameter for DM, in agreement with the
results presented by \citet{Danovich15}. 
This is due to the fact that within the halo the stars are more concentrated
toward the center, whereas the spin of the DM component
is dominated by the outer part of the halo, where most of the angular
momentum of the DM component resides, as also shown in the right panel of Figure \ref{fig:lambda_dmo_dmb+jrm} in Appendix \label{sec:applam}.
In addition, major mergers result in a reduction of the specific angular momentum, as shown in \citet{Welker14}.
On the contrary, the gas component always
has a spin parameter distribution shifted toward larger values,
in agreement with more recent studies by \citet{SS05}, \citet{Kimm11},
\citet{SSBH12}, and \citet{Danovich15}, but in contrast to previous studies by
\citet{vdB02} and \citet{Chen03}, who found nearly the same spin distributions 
for gas and DM. The larger spin values for the gas reflect the
continuous transport of the larger angular momentum from the outer parts
into the center due to gas cooling. While this leads even to a
spinning up of the gas component with time, the distributions for the stars
and DM remain relatively constant. 
This spin-up of the gas component might be caused by the continuously accreted cold gas, which brings in large angular momentum from farther away along the filaments \citep{Pichon11}. 
When dividing the gas into hot and cold phases, we note that the hot gas 
(red) has lower values than the cold gas (blue), which has a long tail to
high $\lambda$-values. This is in contrast to \citet{Chen03},
who found that the spin parameter for the hot gas is higher than that
of the cold gas. On the other hand, our results agree well with more recent
studies by \citet{Danovich15}. This trend for the cold gas to high values was
also seen by \citet{Stewart11}, who have calculated $\lambda$-values for
the cold gas and found values around $0.1 - 0.2$. For a better overview we have listed
the fit value of the $\lambda$-distribution for the different
components in Table \ref{tab:lambda_comp}. 

\begin{table}
\caption{The $\lambda_{0}$ Fit Value for the Different Components at Different Redshifts}
\label{tab:lambda_comp}
\centering
 \begin{tabular}{l l l l l l}
 \hline \hline
  Redshift & $\lambda_\mathrm{stars}$ & $\lambda_\mathrm{DM}$ & $\lambda_\mathrm{gas}$ & $\lambda_\mathrm{hot}$ & $\lambda_\mathrm{cold}$ \\ \hline
  2 & 0.023 & 0.043 & 0.061 & 0.053 & 0.074 \\
  1 & 0.026 & 0.046 & 0.078 & 0.069 & 0.097\\
  0.5 & 0.021 & 0.042 & 0.085 & 0.078 & 0.107\\
  0.1 & 0.020 & 0.042 & 0.090 & 0.083 & 0.123 \\
 \hline  
 \end{tabular}
\end{table}
 
 
\subsection{Alignments}\label{subsec:alignment}

In this subsection we want to study the orientation of the angular momentum vectors of the different components in comparison to each other within the innermost region of the halo. For the calculation of
the innermost region of the halo we only take particles within the
inner 10\% of the virial radius ($R_\mathrm{vir}$) into account, which
corresponds roughly to the size of the galaxies. We chose this radius
as an approximate medium value, since smaller disks have a size of
about 5\% while extended gaseous disks can reach out up to
20\% of the virial radius. 

The angle between two vectors of the species $i$ and $j$ is calculated
by
\begin{equation} \label{eq:angle}
\mathrm{cos}\left(\theta\right) = \frac{\textbf{J}_{i} \cdot \textbf{J}_{j} }
{|\textbf{J}_{i}| \cdot |\textbf{J}_{j}|} .
\end{equation}

We then bin the angles from 0 to 180 degrees in 18 bins of a size of
$10^{\circ}$ each and count the number of halos within each bin. For the
plots the number is normalized to unity.

\begin{figure}
 \begin{centering}
 \includegraphics[trim={-1cm 3.5cm 1cm 5.2cm}, width=0.45\textwidth,
  clip=true]{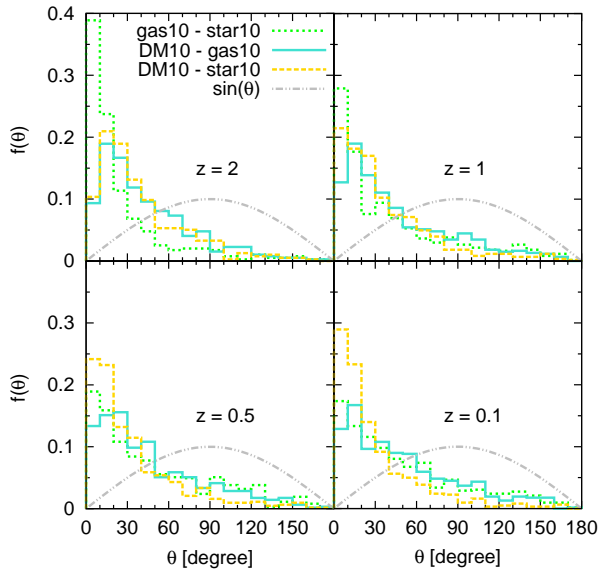}
 \caption{Angle between the total angular momentum vectors of the
  different components within the innermost 10\% of
  $R_\mathrm{vir}$ as indicated in the plots for different
  redshifts. The gray dot-dashed line is expected for a random
  distribution of the angles. At higher redshift the gas and stellar
  components (green dotted) are well aligned, and their alignment
  gets worse with decreasing redshift. In contrast, the alignment of
  the DM and the stars (yellow dashed) gets better with
  decreasing redshift. }
  \label{fig:jangle_10rvir_10rvir}
 \end{centering}
\end{figure}

As shown in Figure \ref{fig:jangle_10rvir_10rvir}, the distribution of the angles between the angular momentum vectors of gas and stars (green
dotted) generally shows a good alignment at higher redshifts and becomes less aligned with decreasing redshift. This is in contrast to the alignment between the DM and stellar components (yellow dashed), which shows a better alignment with decreasing redshift. The angles between the DM and the
gas component (turquoise solid) are an intermediate case, and their distribution does not change with redshift. For comparison, a random distribution 
of alignment angles is shown as a gray dot-dashed line in Figure \ref{fig:jangle_10rvir_10rvir}, demonstrating that none of the angle distributions found in the simulations are of random nature. The difference
between the behavior and especially the evolution of the alignment of
the gas and the stellar component can be explained by the fact that the
angular momentum of the gas reflects the freshly accreted material
(similar to the stars at high redshift), while the stars at low redshift
reflect the overall formation history, similar to the DM content. 
These trends are in qualitative agreement with results obtained from
simulations at very high redshift ($z>9$) by \citet{BiffiMaio2013}, who also
found the angular momentum of the gas component in the center to reflect the recent accretion history. 


\section{Kinematical Split-Up }

In this section we investigate properties of our galaxies as a function
of their stellar mass and angular momentum. We therefore
calculate the angular momentum $\textbf{J}$ of the stars within the
innermost 10\% of the virial radius. Furthermore, we ignore
all particles within the innermost 1\% of the virial radius because of their potentially unknown contributions caused by bulge
components or numerical resolution effects. We rotate the positions
and velocities of all particles such that their $z$-components are
aligned with the angular momentum vector $\textbf{J}$ of the stellar
component. In the same manner we produce another data set that is
rotated such that the $z$-components are aligned with the angular
momentum vector $\textbf{J}$ of the gas component. These data are
used for the $\varepsilon_\mathrm{gas}$-distribution because the gas
circularity calculated with the data rotated according to the angular
momentum of the stars can fail to detect a gas disk that is misaligned
with the stellar disk and vice versa.


\subsection{The Circularity Parameter $\varepsilon$}

\begin{figure}[t!]
\centering
\includegraphics[trim={0.4cm 2.5cm 0.cm 3cm}, clip=true, angle=270,width=0.5\textwidth]{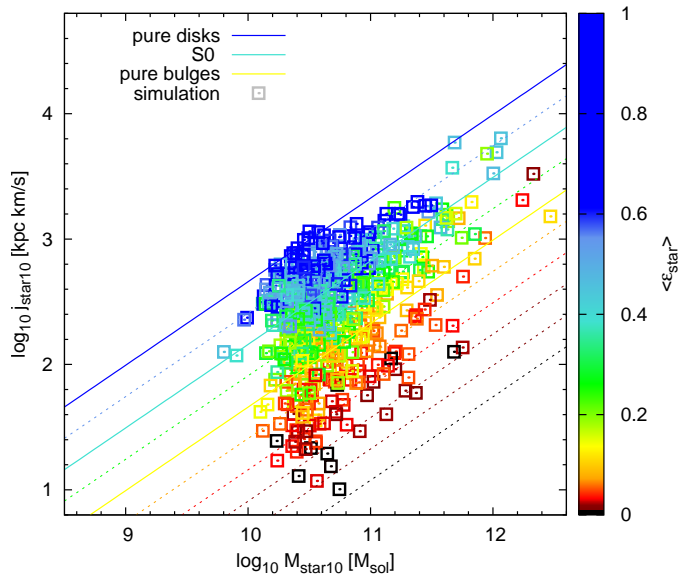}
\caption{Stellar mass in the inner 10\% of the
 virial radius vs. the specific angular momentum of the stars for all
 halos at redshift $z=0.1$. The mean circularity
 $\varepsilon_\mathrm{star}$ of the halos is color-coded.  
 We adopted the scaling relations of \citet{Romanowsky12} (their Figure 2), where the blue line stands for pure disks, the turquoise
 one for S0, and the yellow one for pure bulges. }
\label{fig:jm_eps_av}
\end{figure}

Since in the Magneticum simulations disk and spheroidal galaxies are formed, we need to categorize them for our analysis according to their morphology. 
In order to distinguish between different types of galaxies, we use the
circularity parameter $\varepsilon$. The $\varepsilon$-parameter was
first introduced by \citet{Abadi03} as $\varepsilon_{J} =
J_z/J_\mathrm{circ}(E)$. 
For our study we use the definition of
\cite{Scan08}, which is given by
\begin{equation} \label{eq:circ}
\varepsilon = \frac{j_z}{j_\mathrm{circ}}=
\frac{j_z}{rV_\mathrm{circ}},
\end{equation}
where $j_{z}$ is the $z$-component of the specific angular momentum of
an individual particle and $j_\mathrm{circ}$ is the expected specific
angular momentum of this particle assuming a circular orbit with radius
$r$ around the halo center of mass, with an orbital velocity of
$V_\mathrm{circ}(r)=\sqrt{GM(r)/r}$.

We compute the circularity $\varepsilon$ for every individual particle
between 1\% and 10\% of the virial radius. 
This is done for the stellar and the gas component, where we use
the data that were rotated according to the angular momentum of the
stars or the gas, respectively. 
To obtain the circularity distribution for our selected halos, we compute the fractions $f(\varepsilon)$ of particles within equal-distant bins, using a bin size of $\Delta \varepsilon =0.1$. 
In a dispersion-dominated system there is usually a broad peak in the distribution at $\varepsilon \simeq 0$, while in a
rotation-supported system there is usually a broad peak at $\varepsilon \simeq 1$.

To test the hypothesis that the mass and the angular momentum of galaxies
are the most important ingredients in their formation history and the resulting morphology (e.g., \citealp{Fall83,Romanowsky12,Obreschkow14}), we plot our galaxies at $z=0.1$ in the stellar mass vs. stellar angular momentum plane in Figure \ref{fig:jm_eps_av}. 
Thereto we take \textit{all} star particles within the innermost 10\% into account. 
We color-code the simulated galaxies according to the absolute \footnote{We take the absolute value, since there is a tiny fraction of halos with small $b$-values for which $\varepsilon$ can have a slightly negative value owing to the different weightings in the definitions of the specific angular momentum (eq. \eqref{eq:specangmom}) and the circularity (eq. \eqref{eq:circ}).} value of the mean of their stellar circularity parameter. 
Adopting the classification diagram of galaxies by \citet{Romanowsky12} (their Figure 2), we overplot lines where the specific angular momentum follows the relation $j \propto M^{\alpha}$ with $\alpha \approx 2/3$. 
We plot different lines with distance $\Delta b=0.25$. 
As can clearly be seen in Figure \ref{fig:jm_eps_av}, the mean circularity parameter is following this relation.
Inspired by this result, we classify our galaxies according to what we will in the following refer to as the ``$b$-value'':
\begin{equation}
 b = \mathrm{log}_{10}\left(\frac{j_\mathrm{star}}{\mathrm{kpc}~\mathrm{km/s}}\right) - \frac{2}{3}\mathrm{log}_{10}\left(\frac{M_\mathrm{star}}{M_{\odot}}\right),
\end{equation}
which is the $y$-intercept of the linear relation $f(x)=ax+b$ in the log-log of the stellar-mass--specific-angular-momentum plane.

According to \citet{Romanowsky12}, galaxies with $b$-values close to $-4$ (blue line) are expected to be disks, while $b\approx-5$ indicates pure spheroidals (yellow). 
In our simulations we even find galaxies with $b$-values down to $b=-6.25$ for the galaxies with the smallest specific stellar angular momentum. 

\begin{figure}
\centering
\includegraphics[trim={-0.5cm 3.5cm 1.5cm 5.5cm}, clip=true,width=0.45\textwidth]{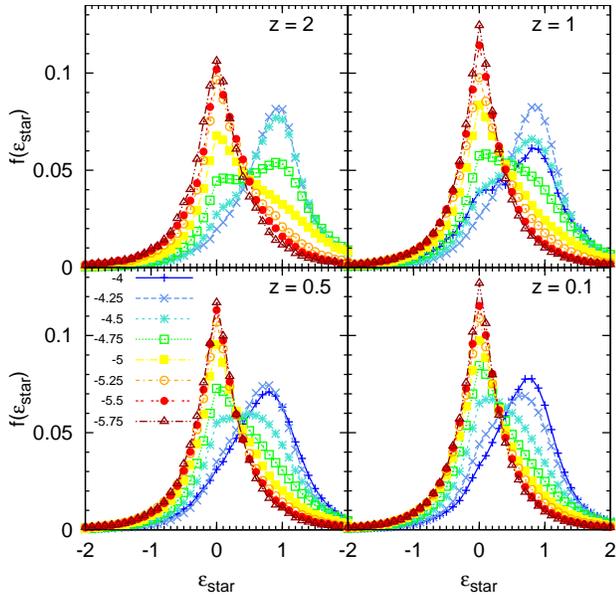}
\caption{Averaged $\varepsilon$-distributions of the
 stellar component for four redshifts. Thereby, each $\varepsilon$
 bin is averaged over all halos that lie in the corresponding
 $b$-value bin. We see a clear transition of the galaxy types from rotational-supported systems to dispersion-dominated systems
 at all redshifts.}
\label{fig:av_eps_bin}
\end{figure}

Figure \ref{fig:av_eps_bin} shows the averaged
$\varepsilon$-distribution of the stellar component at four
redshifts, colored according to the different $b$-value bins shown in Figure \ref{fig:jm_eps_av}. 
Each $\varepsilon$ bin is averaged over the halos that reside in the
chosen $b$-value bin. 
There is a clear transition between the galaxies with different dynamics, from rotational-supported ($\langle \varepsilon \rangle = 1$) to dispersion-supported ($\langle \varepsilon \rangle = 0$) systems, reflected by their $b$-value from the
$M_\mathrm{star}$-$j_\mathrm{star}$-plane: 
While galaxies with $b\approx-4$ clearly have most stars around $\varepsilon = 1$, and thus are dominated by rotation, galaxies with $b\approx -6$ peak around $\varepsilon = 0$, indicating that there is no significant rotation. This is true for all redshifts. However, we see a slight trend with redshift at intermediate $b$-values: 
for example, the distribution of the halos in a $b$-value bin of
$-4.75$ (green squares) at $z=2$ has the higher of the two peaks
around $\varepsilon=1$, whereas at $z=0.1$ it only peaks around
$\varepsilon=0$. It is also interesting that the galaxies in the
$b$-value bin of $-4.5$ (turquoise stars) at redshift $z=2$ have a
dominant rotationally ordered component, which becomes more dispersion
dominated with decreasing redshift. We also note that the
distributions around $\varepsilon=0$ become slightly more distinct at lower
redshifts. This indicates that there is an evolution of the
different galaxy types with cosmic time along the $M_\mathrm{star}$-$j_\mathrm{star}$-plane.

\begin{figure}
\centering
\includegraphics[trim={0.cm -0.1cm 0.5cm 0cm}, clip=true,width=0.95\columnwidth]{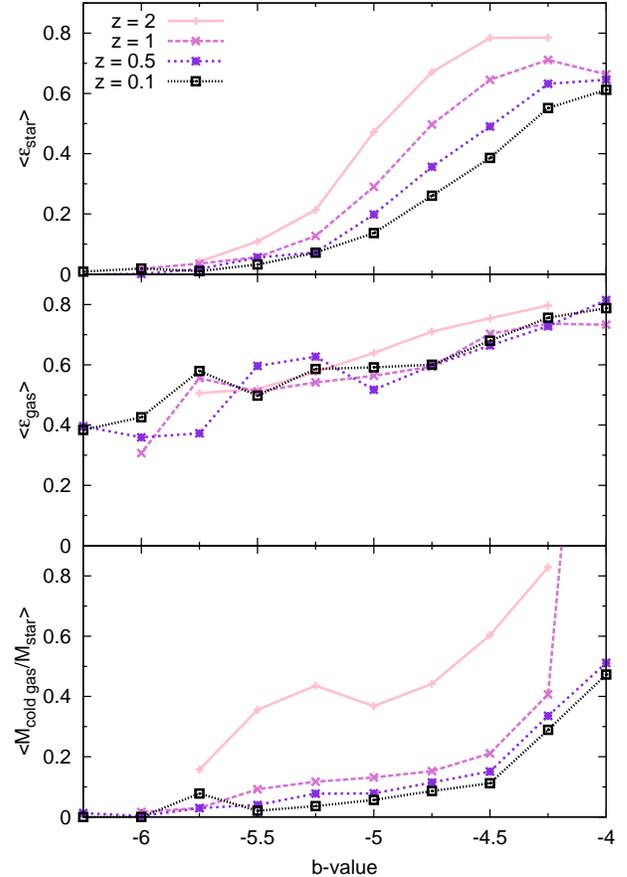}
\caption{Top: mean circularity $\varepsilon$ of the star component averaged over all halos in the corresponding $b$-value bin. 
 There is a clear trend for decreasing
 mean $\varepsilon$ when moving from the pure disks (high $b$-value)
 down to the pure bulges (smaller $b$-value) for all four redshifts.
 Middle: same as the top panel, but for the $\varepsilon$ of the gas component.
 Bottom: The mass fraction of cold gas with respect to the stars,
 both within the inner 10\% of the virial radius, averaged
 over all halos in the corresponding bin. At high redshift (pink
 solid line) there is much more cold gas, even in halos with a smaller
 $b$-value. }
\label{fig:jm_av_bin}
\end{figure}

This additional evolutionary trend gets more evident in Figure \ref{fig:jm_av_bin},
which shows $\varepsilon_\mathrm{star}$ (top),
$\varepsilon_\mathrm{gas}$ (middle), and the mass fraction of the cold
gas with respect to the stellar mass (bottom) versus the $b$-value, with spheroidal systems at the left and disk-like systems on the right. 
Thereby, each of these properties is averaged over the halos in the corresponding $b$-value
bin. We can clearly see that the average
$\varepsilon_\mathrm{star}$ increases with increasing $b$-value at all four redshifts. 
Additionally, the average stellar circularity is generally larger at higher redshifts for all types of galaxies, albeit this trend is stronger for galaxies with larger $b$-values. 
This clearly shows that disk galaxies at higher redshifts had less prominent central bulges that would shift the value toward $\langle \varepsilon \rangle = 0$. 
For $\langle\varepsilon_\mathrm{gas}\rangle$ we do not find any clear trend. 
The bottom panel of Figure \ref{fig:jm_av_bin} shows a similar trend for the mean cold gas fraction. 
The spheroidal systems at present day with small $b$-values have only low amounts of gas, compared to the stellar mass, usually below 10\%, whereas disks have gas fractions of 20\% or more. For higher redshifts this fraction increases successively, and the strongest evolution is visible
between redshifts $z=2$ (pink solid line) and $z=1$. Below $z=1$ there is
only a mild but continuous\footnote{At $z=1$ (magenta dashed line) the value for the galaxies with the highest $b$-value exceeds the scale, since there are six galaxies that on average have a cold gas fraction of 2.53 with respect to the stars.} evolution, but at $z=2$ even spheroidals have gas fractions of 20\% or more, while disk galaxies can contain more than 40\% gas compared to their stellar content. For the largest $b$-values the galaxies can even be dominated by gas, i.e., the gas fraction is larger than 50\%.  
Thus, we conclude also that galaxies that are spheroidal systems have a nonnegligible
cold gas fraction at high redshifts.


\subsection{The $\lambda$-parameter}\label{subsec:lambda}

\begin{figure*}
 \begin{centering}
  \includegraphics[trim={2cm 0cm 1.5cm -0.5cm},
   width=0.55\textwidth, clip=true,
   angle=270]{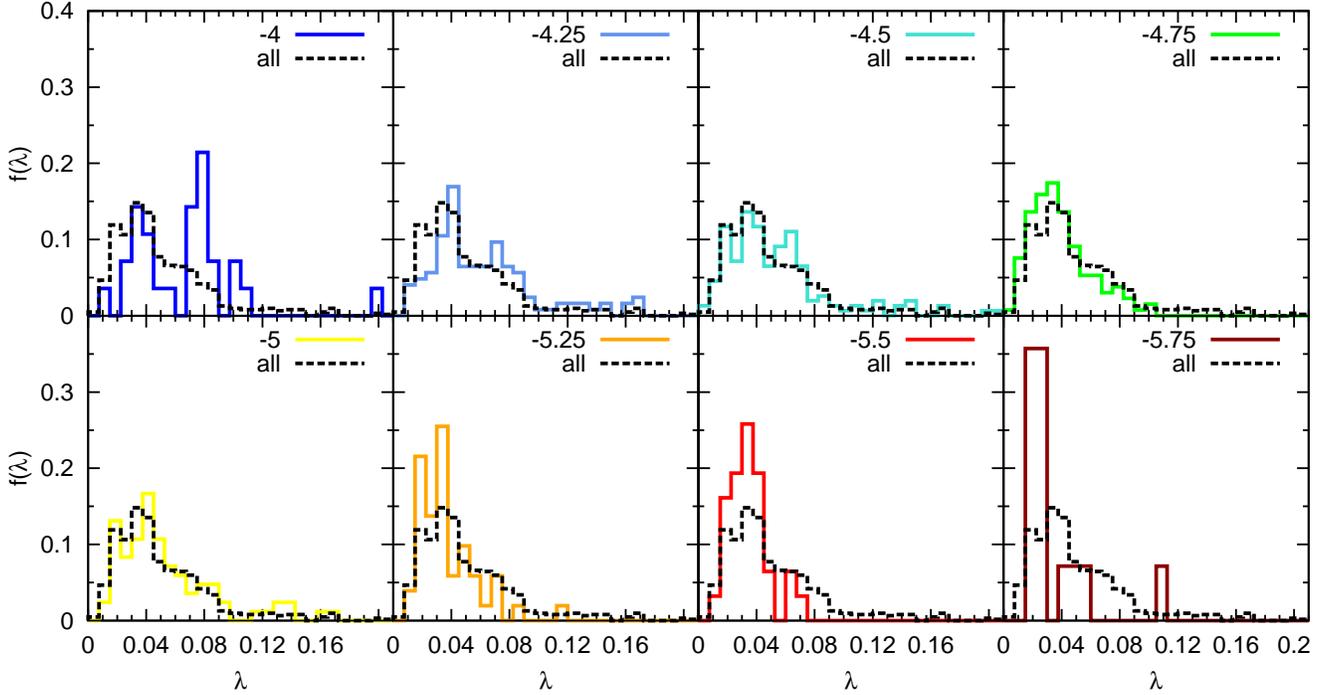}
  \caption{Distribution of the total $\lambda$-parameter for
   the halos according to their $b$-values at $z=0.1$. There is a
   transition from the disks (upper left) to the bulges (lower
   right), i.e., from rotation-dominated systems (higher
   $\lambda$-values) to dispersion-dominated systems (lower
   $\lambda$-values). Interestingly, the halos with the highest
   $b$-values show a dichotomy in their distribution.
  }
  \label{fig:lambda_bv}
 \end{centering}
\end{figure*}

We want to understand whether this classification of the galaxies according to their position on the $M_\mathrm{star}$-$j_\mathrm{star}$-plane has an effect on the spin parameter $\lambda$.
In Figure \ref{fig:lambda_bv} we show the distribution of the total
$\lambda$-parameter of the whole halo content (see
Equation \ref{eq:lambda}), exemplary for redshift $z=0.1$. We divide the
halos according to their $b$-value. The first interesting
thing to note is that there is a double peak in the
$\lambda$-distribution for galaxies with disk-like kinematics (upper left panel). 
This could indicate that there are different formation channels for disk-like systems, e.g., gas-rich major mergers, as discussed in \citet{Springel2005mm} and \citet{schlacht:1}.
The peak at higher $\lambda$-values becomes smaller
as we move to the right and seems to disappear at a $b$-value of
$-4.75$. Galaxies with $b$-values below $-5.25$ are hosted in halos
with significantly smaller $\lambda$-values compared to the overall
distribution (black dashed line). Therefore, on average there seems
to be a connection between the galaxy types and the spin parameter $\lambda$ of the hosting halos,
reflected in a continuous transition of the spin parameter distribution of the
host halo with the $b$-value obtained from the $M_\mathrm{star}$-$j_\mathrm{star}$-plane.


\subsection{Alignments of the Central Components}\label{subsec:alignment}

\begin{figure*}
 \begin{centering}
  \includegraphics[trim={2cm 0cm 1.5cm -0.5cm},
   width=0.55\textwidth, clip=true,
   angle=270]{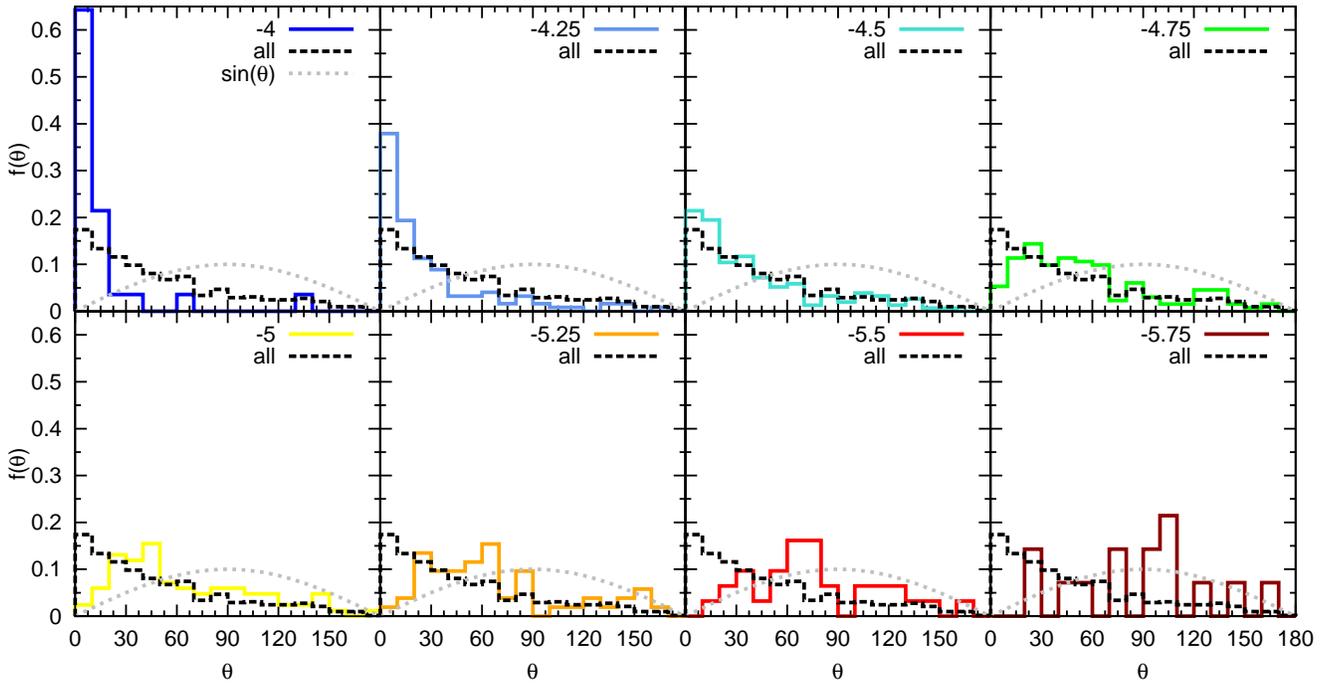}
  \caption{Angle between the angular momentum vector of the gas
   and that of the stars within the innermost 10\% of the
   virial radius according to their $b$-values at redshift $z=0.1$. There is a clear transition between the galaxy
   types. The higher the $b$-value, the better aligned the stellar and
   gas components are. When going to lower $b$-values, i.e., moving down
   on the $M_\mathrm{star}$-$j_\mathrm{star}$-plane to the bulges,
   the angles of the two components seem to be randomly distributed
   (compare gray dotted sinusoidal distribution). }
  \label{fig:jgasstarangle_bvalue}
 \end{centering}
\end{figure*}

To verify the dynamical connection between the different baryonic
components within the galaxies, we show in Figure \ref{fig:jgasstarangle_bvalue}
the distributions of the alignment between the central angular momentum of the
stellar and the gas component. As before, we classify our galaxies according to the $b$-value.
By moving from high (upper left) to low (lower right) $b$-values we
immediately see that the alignment between the angles is good for the disk galaxies (large $b$-values) and gets increasingly more random for decreasing $b$-values. 
This is indicated by the gray dotted line, which again illustrates a random distribution (as in Figure \ref{fig:jangle_10rvir_10rvir}).
We find a continuous transition from the disk-like to the
bulge-dominated galaxies. It illustrates that in spheroidal
systems the stars are significantly misaligned with an eventually
present gaseous disk. This result is supported by the findings of the
$\mathrm{ATLAS^{3D}}$-project \citep{Cappellari11} that a
kinematical misalignment of the gas component with respect to the
stars is not unusual \citep{Davis11}. 


\section{Classification of Simulated Galaxies}

So far we have seen that there is a continuous transition between the
different types of galaxies within the $M_\mathrm{star}$-$j_\mathrm{star}$-plane,
orthogonal to $j \propto M^{\alpha}$ with $\alpha \approx 2/3$.
We now want to study the kinematical properties of our simulated galaxies depending on their classification as disk and spheroidal galaxies. 


\subsection{Selection Criteria}

In our simulations, many spheroidal galaxies show extended, ring-like structures of cold gas, in good agreement with recent observations 
\citep{Salim2010}. Therefore, including the circularity of the gas
within 10\% of the virial radius as a tracer of morphology can lead to a misinterpretation. Additionally, there can be huge 
uncertainties for galaxies that have almost no gas left.
Using a criterion only based on the $b$-value of the galaxies also
is not straightforward because of the existence of fast rotators
among the spheroidal galaxies. 
Observationally there seems to exist some overlap between the different galaxy types within
the $M_\mathrm{star}$-$j_\mathrm{star}$-plane. In principal,
the luminosity of the stars could be taken into account,
since old stars that make up the bulge and the halo stars are not as
luminous as young stars that build up the disk. Hence, in
observations a galaxy with an old massive stellar bulge and an
extended disk of young stars and gas is very likely to be
classified as a spiral galaxy. 

However, here we will stick to a classification of galaxies
based on the circularity distribution (Equation \ref{eq:circ}) of their stars,
which allows us to capture rotationally supported stellar disk structures or
dispersion-dominated spheroidal structures.
We combine this criterion with the mass fraction of the cold gas with respect to the stars, following our result from Figure \ref{fig:jm_av_bin}. 
As before, circularity distributions are evaluated within 10\% of the virial radius, excluding the central
1\%, while we include the central 1\% in our calculation of the cold gas fraction as the resolution is not important in this case.
We use the results from the previous section to justify the threshold values
used to select dispersion-dominated, gas-poor spheroidal and rotational-supported,
gas-rich disk galaxies to reflect counterparts of classical, observed 
elliptical, and spiral galaxies.

As shown in Figure \ref{fig:av_eps_bin}, there is a clear, bimodal
behavior of the epsilon distribution within different $b$-value bins. 
We show in Appendix \label{sec:appclass} in detail the cumulative distributions that allow us to define the proper thresholds bracketing the transition
regions. These thresholds are then applied to the circularity distributions
of the individual galaxies, which of course in general show a more
complex shape than the averaged distributions. 
The left panel of Figure \ref{fig:mfrac+alignment} in Appendix \label{sec:appclass} shows the cold gas mass fraction of the galaxies classified only by their circularity $\varepsilon_\mathrm{star}$, illustrating our choice of including a gas criterion. 
In short, we are using the following selection criteria: 
\begin{itemize}
\item We classify a galaxy as a spheroidal galaxy if the majority of
 particles is in a close interval around the origin (i.e., $f(-0.4
 \leq \varepsilon \leq 0.4) \geq 0.6$). This percentage cut is
 adapted to the redshift. For the details see the left panels of
 Figure \ref{fig:cum_eps} in Appendix \label{sec:appclass}. 
 
 In addition, it has to
 have a cold gas mass fraction with respect to the stars lower than
 $0.35$ at $z=2$, $0.2$ at $z=1$, $0.125$ at $z=0.5$, and $0.065$ at
 $z=0.1$. This criterion is chosen such that the fraction values
 satisfy the linear function $f(z)=0.15 \cdot z + 0.05$. 
 The upper limit for the spheroidal galaxies at redshift $z=0.1$ 
 seems plausible, since \citet{Young14} found that massive elliptical
 galaxies (of the red sequence) have mass fractions of $HI$ and $H_2$, 
 compared to the stars, up to 6\% and 1\%, respectively. 
 
\item We classify a galaxy as a disk galaxy if the majority of
 particles are off-centered from the origin (i.e., $f(0.7 \leq
 \varepsilon \leq 3) \geq 0.4$; for the dependence on the redshift see
 right panels of Figure \ref{fig:cum_eps} in Appendix \label{sec:appclass}). 
 
 Additionally, the constraint on the mass fraction of
 the cold gas is such that it has to be higher than $0.5$ at $z=2$,
 $0.35$ at $z=1$, $0.275$ at $z=0.5$, and $0.215$ at $z=0.1$. This
 criterion is chosen such that the values of the mass fraction
 satisfy the linear function $f(z)=0.15 \cdot z + 0.2$.
 
\item All remaining galaxies, which fulfill neither of the two
 criteria, are classified as ``others.'' 
\end{itemize}

The total number of galaxies above the mass cut of $5 \cdot
10^{11}M_{\odot}$ selected from our simulation at redshift $z=0.1$ is
622 (for other redshifts see Table \ref{tab:halono}). According to
the above criteria, 64 galaxies are classified as classical disks and 110 as
classical spheroids, while 448 are classified as ``others.''
Such unclassified galaxies include merging objects, bulge-dominated spirals,
spheroids with extended gas disks, barred galaxies, or irregular objects.

\begin{table}[h!]
\caption{Number of Halos in the Selected Mass Range} 
\label{tab:halono}
\centering
\begin{tabular}{c c c c c c }
  \hline \hline 
  Redshift & $N$ & $N_\mathrm{spheroid}$ & $N_\mathrm{disk}$
  \\ \hline 
  2 & 396 & 34 & 89   \\ 
  1 & 606 & 87 & 73   \\ 
  0.5 & 629 & 146 & 59   \\ 
  0.1 & 622 & 110 & 64 \\ 
  \hline   
\end{tabular}
\tablecomments{Total number of halos $N$ in the selected mass range, the
  number of spheroids $N_\mathrm{spheroid}$ and disks
  $N_\mathrm{disk}$ (classified by the circularity of the stellar
  component and the mass fraction of the cold gas with respect to the stars)
  at different redshifts.} 
\end{table}

In order to identify parts of the unclassified objects, the
classification could be refined in such a way that we additionally
divide into spheroids that have a gaseous star-forming disk or galaxies that have a large stellar disk with only little gas (S0 galaxies). 
At lower redshifts our classification becomes increasingly difficult since there are a lot of
galaxies that have a very dominant stellar bulge but, on the other
hand, also possess an extended gaseous disk containing many young
stars. 

Although the criterion for the gas mass fraction is relatively arbitrary
and may overestimate the number of spheroids, we checked
that changing this selection criterion does not change the results
qualitatively, although adding/removing galaxies from the spheroidal sample.
At higher redshifts an additional difficulty might be that the transition
of galaxy types is less strict, as seen in Figure \ref{fig:av_eps_bin}. At these
redshifts, remaining dynamical signatures of the current formation process
will be more pronounced and individual distributions of the circularity
parameter might be more complex. Thus, a clear assignment might fail.


\subsection{Comparison of the Simulated Stellar Specific Angular Momentum with Observations}\label{def:specamstar}

\begin{figure*}
  \begin{centering}
    \includegraphics[height=7cm,clip=true]{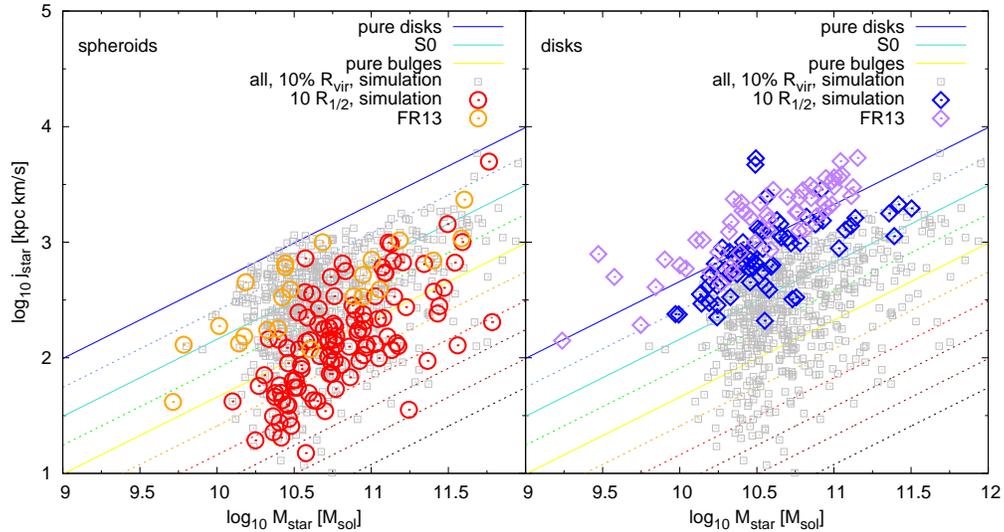}
    \caption{Galaxies within the stellar-mass--specific-angular-momentum plane at $z=0.1$. 
    Gray symbols represent all galaxies extracted from the simulations (as in Figure \ref{fig:jm_eps_av}), where angular momentum and stellar mass are measured within 10\% of the virial radius. 
    The left panel shows the simulated galaxies classified as spheroids (red circles) compared to data from observed elliptical galaxies by \citet{Fall13} (orange circles, denoted by ``FR13''). The right panel shows disks in our simulation (blue diamonds) compared to observational data of spiral galaxies by FR13 (purple diamonds). For the disks and the spheroids, we evaluated the mass and the specific angular momentum within $10 ~ R_{1/2}$, which more closely resembles the radii studied by FR13. The colored lines correspond to the $b$-values (see Figure \ref{fig:jm_eps_av}). 
    }
    \label{fig:jstarmass}
  \end{centering}
\end{figure*}

After having extracted a set of spheroidal and disk galaxies, we can
evaluate the relation between the specific angular momentum and the
mass of the stellar component and compare them with observations, as shown in Figure \ref{fig:jstarmass}.
To compare with observations presented by \citet{Fall13} (hereafter FR13) we calculate the mass and specific angular momentum for all stellar particles within $10 ~ R_{1/2}$ instead of $10 \% R_\mathrm{vir}$, where $R_{1/2}$ is the radius that contains half the stellar mass of the galaxy and roughly corresponds to observed effective radii. 
This is done to account for the fact that, for a given mass, disk galaxies have a larger effective radius than spheroids \citep{Shen03}, and to better resemble the radius ranges studied in FR13.
The left panel shows the $M_\mathrm{star}$-$j_\mathrm{star}$-plane for our galaxies classified as spheroids (red circles), including the 23 elliptical galaxies (orange circles) presented in FR13 (their Figure 2).
In the right panel the same is shown for our disk galaxies (blue diamonds) in comparison to the 57 spiral galaxies (purple diamonds) from FR13. 
In general, our simulated galaxies are in good agreement with the observations. 
This result has already been shown for a subset of our galaxies at $z=0$ with a rather crude classification criterion in \citet{Remus15}, and we find an even better agreement with observations with our more advanced classification scheme. 
Recently, \citet{Genel15} have shown a comparison of the galaxies in the Illustris simulation with those observations, which are also in good agreement when a similar feedback mechanism is used. 

To understand the impact of the choice of radius used to calculate the angular momentum and the stellar mass, we included the values for all galaxies evaluated within 10\% of the virial radius, as previously shown in Figure \ref{fig:jm_eps_av}. 
For the spheroidal galaxies the effect is much stronger, and we clearly see that the angular momentum is larger for larger radii. 
The radius dependence is shown in more detail in the right panel of Figure \ref{fig:lambda_dmo_dmb+jrm} in Appendix \label{sec:applam}, where we show that the angular momentum of spheroids increases with radius and that there is a significant contribution to the specific angular momentum from large radii. 
For disk galaxies the exact radius is less important, as most of the angular momentum is contained within the disk at relatively small radii. 
The same holds true for the $b$-value, which more strongly depends on the considered radius for spheroids than for disks, which can be seen in Figure \ref{fig:jrm-bv} in Appendix \label{sec:apprsap}. 

Note that we compare here the specific angular momentum directly measured from the total stellar component from the simulations (Equation \ref{eq:specangmom}), while the observations are inferred from the projected measurements. 
We also do not resolve galaxies with stellar masses smaller than $\approx 10^{10} M_{\odot}$, while the observations include some objects with smaller masses. 


\subsection{The Gas and Stellar Specific Angular Momentum in Simulations and Observations}\label{def:specamgas}

We also compare the specific angular momentum of the gas with that of the stars for all classified galaxies in our simulations. 
\begin{figure}
\centering \includegraphics[trim={0cm 3.5cm 1.5cm 5.2cm},
  width=0.45\textwidth,clip=true]{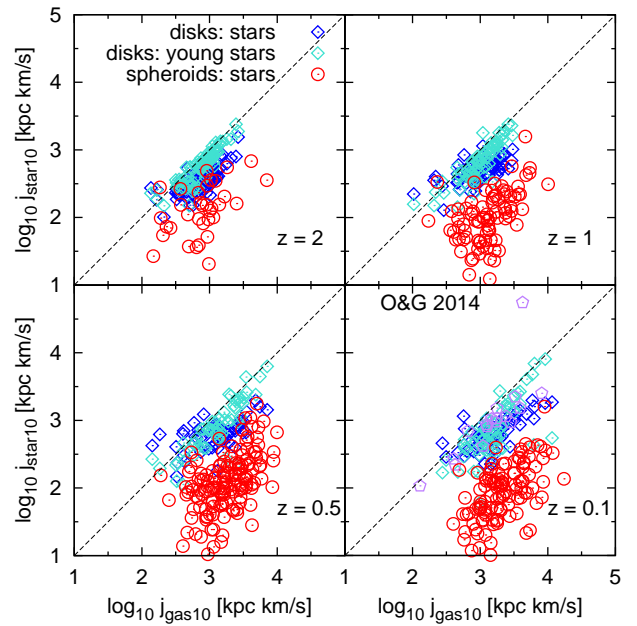}
\caption{Specific angular momentum of the gas against the specific
  angular momentum of stars, both within 10\% of the virial
  radius for galaxies which are classified as disks (blue diamonds) at
  four redshifts as indicated in the plots. Additionally, we show the
  relation also considering only young stars (turquoise diamonds). At
  $z=0.1$ we overplot observational data points calculated by
  \citet{Obreschkow14} (purple pentagons). The values agree well with
  the observations, and we note an overall spin-up with cosmic time.}
\label{fig:jstargas}
\end{figure}
Figure \ref{fig:jstargas} shows the relations for all four redshifts as
indicated in the plots. At redshift $z=0.1$ we include the
observational data taken from the THINGS \citep[The HI Nearby Galaxy Survey,][]{THINGS} sample, which consists of 16 spiral galaxies in the local Universe, for which
surface densities of stars and cold gas are available \citep{Leroy08}.
Specific angular momenta for the gas and stellar components of those galaxies were presented by \citet{Obreschkow14}.

For the disk galaxies, we find that the specific angular momentum of all stars in the central region is slightly smaller than that of the gas (blue diamonds), in agreement with the observations (purple symbols).
This, most likely, originates from the fact that the specific angular momentum of the gas is constantly replenished by freshly accreted gas, which transports larger angular momentum from
the outer parts of the halo to the center.
This becomes more evident by looking at the newly formed stars (turquoise diamonds) in our simulations. 
We consider a star to be young if its formation happened not more than $\Delta z = 0.05 \times 1/(1+z)$ ago at a given redshift $z$, which corresponds to a stellar age of $0.2 \mathrm{Gyr}$ at $z=2$, $0.4 \mathrm{Gyr}$ at $z=1$, $0.5 \mathrm{Gyr}$ at $z=0.5$, and $0.7 \mathrm{Gyr}$ at $z=0.1$. 
If we only take the young stars into account, we find almost an equality (dotted line) with the specific angular momentum of the gas. 
At a lower redshift (lower right panel) the values of the specific
angular momentum of the gas are slightly larger than at higher
redshift (upper left panel). This behavior reflects the spin-up of
the cold gas with time, as already seen before. Hence, also the young stars have
higher specific angular momentum. This appears in a slight trend for
a separation of the young stars and the total stellar component, which
have lower specific angular momentum than the gas. 

The spheroids (red circles in Figure \ref{fig:jstargas}), however, have a significantly lower specific stellar angular momentum compared to that of their gas. 
This suggests that, especially at high $z$, most gas in spheroids originates from the accretion of material from large radii (cold streams or infalling substructures) and hence transports
the higher angular momentum from the outer parts into the center. 

We can clearly see that the spheroids and the disk galaxies show a different
behavior in the relation between the angular momentum of the gas and stars. We
also have seen that the gas gains angular momentum over time, especially in disk
galaxies. In disk galaxies, stars have slightly smaller specific angular momentum
than the gas, and they also show a mild difference between the specific angular
momentum of stellar components with different ages. Here the young stars have
slightly larger specific angular momentum, basically reflecting the angular momentum of the gas from which they form.

All in all, the results from the simulations fit well with the
observational data. In particular, we are in agreement with
\citet{Fall83}, who find that the specific angular momentum of the
galaxies increases with the disk-to-bulge ratio for a given mass. 

We note that the evaluation of this quantity for spheroids especially
at lower redshifts might be fault-prone since, owing to the selection criterion,
there is only a small amount of cold gas within the chosen radial range.


\subsection{Comparison of the Gas and DM Specific Angular Momenta}\label{def:specamgasdm}

Finally, we compare the specific angular momenta of DM and gas.
In particular, we are interested in the scaling relations of the
corresponding individual specific angular momenta of baryonic and
nonbaryonic matter. Previous studies (e.g., \citealp{Fall83,MMW98}) suggest an equality between the angular momentum of the total DM halo and that of the central gas component of disk galaxies.

\begin{figure}
\begin{centering}
  \includegraphics[width=0.42\textwidth, clip=true]{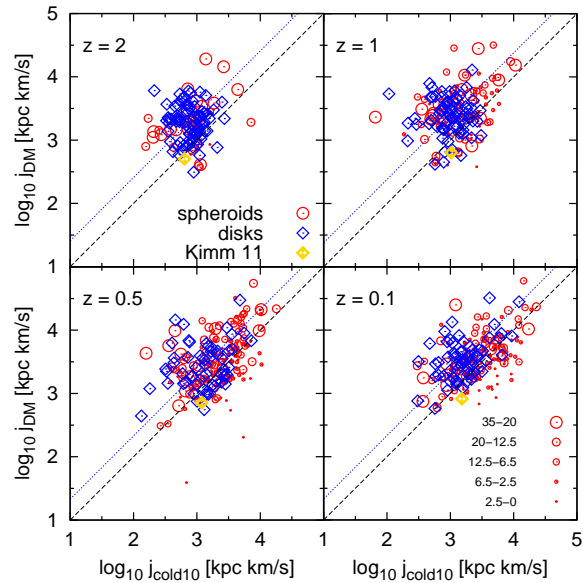}
  \caption{Specific angular momentum of the cold gas within 10\% of the virial radius against the specific angular
    momentum of DM within the entire virial radius. The blue
    diamonds show the disks and the red circles the spheroidal
    galaxies, where the size of the circles reflects the fraction of
    the cold gas mass with respect to the stellar mass. The blue
    dotted line shows the fit for the disk galaxies. The 1:1 relation
    is represented by the black dashed line. For comparison, we add the results obtained by \citet{Kimm11} (yellow diamonds).}
\label{fig:jdmgas10}
\end{centering}
\end{figure}

The relation between the specific angular momentum of the cold gas of
the galaxy, which resides within the innermost 10\% of the
virial radius, and that of the total DM halo is shown in
Figure \ref{fig:jdmgas10} for different redshifts. In general, we find
the specific angular momentum of the DM to exceed (by a
factor of $\approx2$) the specific angular momentum of the cold
gas. Interestingly, the cold gas in disk and spheroidal galaxies does
behave in the same way. This is due to the fact that cold gas always
settles in disk-like structures, even in elliptical galaxies
(e.g., \citealp{Lees91,Young11}). We plot all disks above the
redshift-dependent $M_\mathrm{cold}/M_\mathrm{star}$-cut (blue
diamonds). Additionally, we fit a line parallel to the 1:1 relation
(blue dotted) to the data, in order to see the offset. For all shown
redshifts we find values of the $y$-intercept between $0.33$ and $0.41$
in log space, which is between $2.14$~kpc km/s and $2.57$~kpc km/s in
normal space. However, the scatter around this fit curve is very
large. 
The yellow diamonds show the value for one simulated galaxy  at different redshifts as presented in \citet{Kimm11}. 
Their disk galaxy has only slightly higher specific angular momentum in the gas component compared to the DM halo, but sits within the distribution of values we find in our simulation for different galaxies. 
Our values are also broadly in line with the results of \citet{Danovich15}, who find that the spin of the disk is comparable to that of the DM halo.
The size of the symbols of the
spheroids (red circles) is plotted according to their fraction of cold
gas mass with respect to the stellar mass.  For completeness we plot
all spheroids with a cold gas mass fraction lower than 35\% for all redshifts.  At higher redshift most spheroids have a
large amount of cold gas.  At the lowest redshift there are still some
objects that were classified as spheroids with the
$\varepsilon_\mathrm{star}$-criterion but have a high cold gas mass
fraction.  Interestingly, there are some spheroids with high specific
cold gas angular momentum, which might be due to individual infalling
clumps and small substructures, as these are spheroidal galaxies that
have extremely small cold gas fractions.  However, in such cases we
cannot relate the specific angular momentum of these individual
structures to the specific angular momentum of the halo.

\subsection{Misalignment Angles}\label{subsec:alignment}

As seen before, another indicator for the different evolutionary
states reflected by the stars, DM, and gas are the misalignment
angles between these three components. We now investigate their behavior by focusing on our selected disk and spheroidal galaxies. 


\subsubsection{The Angle between Gas and Stars}

\begin{figure}
 \begin{centering}
 \includegraphics[width=0.44\textwidth, clip=true]{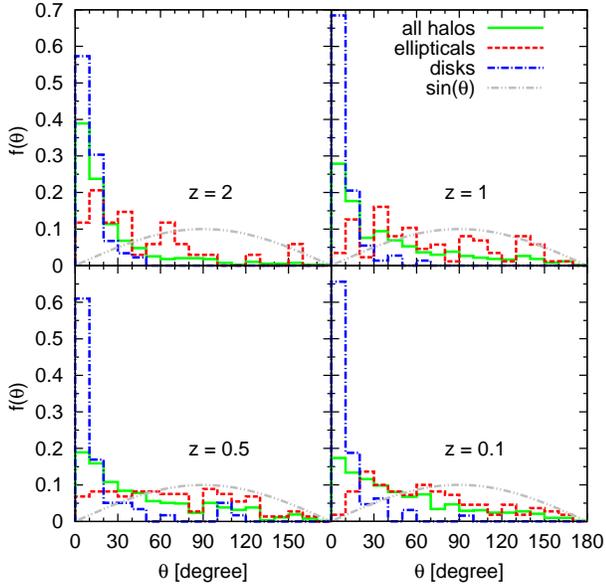}
 \caption{Angle between the angular momentum vector of gas and
  stars, both within the innermost 10\% of the virial
  radius. At all redshifts the disks (blue dot-dashed) are well
  aligned. In contrast, the spheroids (red dashed) are randomly
  distributed. The overall distribution (green solid) gets worse
  aligned with decreasing redshift. }
 \label{fig:jgasstarangle}
 \end{centering}
\end{figure}

At first, we investigate whether we find a correlation between the morphological type and the angles between the angular momenta of the gas and the
stars, both within the innermost 10\% of the virial radius. 
Figure \ref{fig:jgasstarangle} shows the
angle for four different redshifts. 
The disk galaxies (blue dot-dashed histograms) have very well aligned gas and stellar angular momentum vectors with median values of $7.8^{\circ}$ at redshifts $z=2$ and $z=0.1$ and median values of $6.3^{\circ}$ and $7.6^{\circ}$ at redshifts $z=1$ and $z=0.5$, respectively. 
This is in good agreement with \citet{Hahn10}, who found median angles of about $8^{\circ}$ for their disks at $z=1$ and a median of $7^{\circ}$ at redshifts $z=0.5$ and $z=0$. 
It demonstrates that the stellar and gaseous disks in disk galaxies are very well aligned, which is what we expect, since the stars form out of the gas and thus maintain the same orientation. 

The spheroids have a random distribution with a median value of $60.4^{\circ}$, i.e., the gas and stellar components in spheroidal galaxies are often misaligned. 
In summary, the gas and star components of spheroids become less aligned with decreasing redshift, while there is no change for disk galaxies (see Table \ref{tab:angles_g10_s10}). 
The overall trend for all halos in the simulation shows the same behavior as the spheroids, in agreement with Figure \ref{fig:jgasstarangle_bvalue}.

\begin{table}
\caption{Median Misalignment Angles between the Baryons in the Center}
\label{tab:angles_g10_s10}
\centering
 \begin{tabular}{l c c c }
 \hline \hline
  Redshift & All Halos & Disks & Spheroids \\ \hline
  2 & 13.4 & 7.8 & 37.4\\ 
  1 & 24.4 & 6.3 & 56.5\\
  0.5 & 35.8 & 7.6 & 65.5 \\
  0.1 & 37.6 & 7.8 & 60.4 \\
 \hline  
 \end{tabular}
\tablecomments{The median misalignment angles between the angular momentum vectors of gas and stars, both within the inner 10\% of the virial radius ($\theta_\mathrm{gas10-stars10}$).} 
\end{table}


\subsubsection{The Angle between DM and Baryons}

\begin{figure*}
 \begin{centering}
 \includegraphics[width=0.9\textwidth, clip=true]{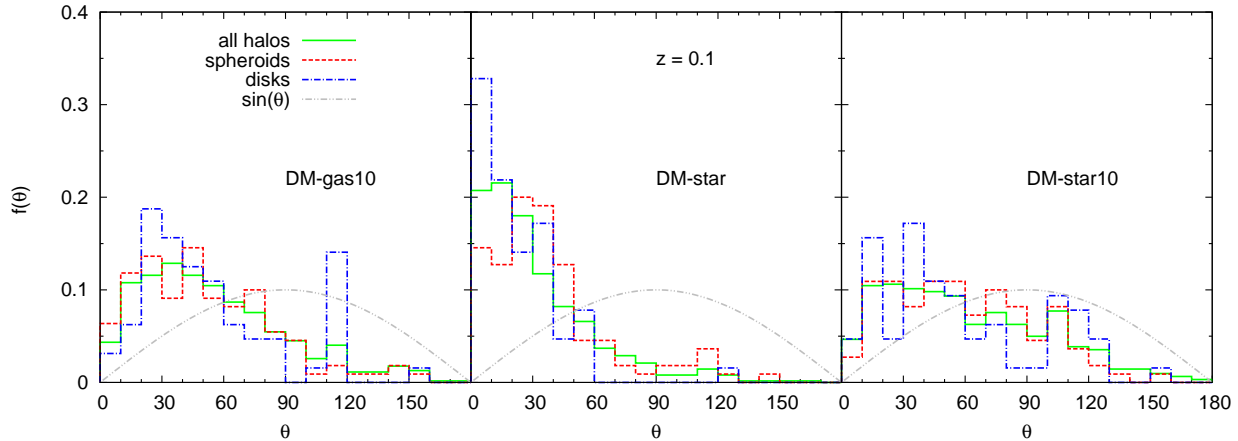}
 \caption{Left: angle between the total angular momentum vector
  of DM in $R_\mathrm{vir}$ and the gas within the innermost 10\% of $R_\mathrm{vir}$. The overall alignment is poor. 
  Middle: The angle between
  the total angular momentum vectors of the DM and stars
  within $R_\mathrm{vir}$. The disk galaxies (blue dot-dashed
  lines) seem significantly better aligned than the spheroids (red dot-dashed
  lines). Right: angle between the total angular momentum
  vector of the DM in $R_\mathrm{vir}$ and the stars within
  the innermost 10\% of the virial radius. For all shown
  distributions the alignment is very poor, almost random. }
  \label{fig:jdmgasstarangle}
 \end{centering}
\end{figure*}

Since we see a clear difference in the alignment between the angular momenta of the stars and gas for spheroids and disks, we want to test whether this is reflected in the relation between the angular momentum of the baryonic components and DM. 
The left panel of Figure \ref{fig:jdmgasstarangle} shows the misalignment angle between
the angular momentum of the total DM halo and the gas within
the inner 10\% of the virial radius at redshift $z=0.1$.
Again, the gray dashed line is the expected distribution if the angles were spread randomly. 
We clearly see that the median misalignment found for all halos (green solid lines) is $\approx49^{\circ}$, for the disks $\approx45^{\circ}$, and for the spheroids $\approx47^{\circ}$. This is slightly larger than the results of 
\citet{SSBH12}, who found a median misalignment angle of about $30^{\circ}$
for the gas within the innermost 10\% of the virial
radius compared to the total angular momentum vector of all halos.
On the other hand, \citet{Hahn10} compared the angular momentum of the
gas component of the disk to the total angular momentum and obtained
about $49^{\circ}$ at $z=0$, well in line with our results. 

We now want to see under which circumstances the orientation of the angular
momentum vectors of the total DM halo is reflected in that of 
the stellar component, since we have seen in Figure \ref{fig:jgasstarangle}
that the angular momentum vectors of the gas and that of the stars are very
well aligned in the inner part of the halo for disk galaxies and poorly aligned
for spheroids. 
The angle between the angular momentum vectors of the DM and that of
the stars within the virial radius is shown in the middle panel of
Figure \ref{fig:jdmgasstarangle}. In general, there is an alignment of the two
components. The disk galaxies are slightly better
aligned with a median angle of $\approx18^{\circ}$ compared to the spheroids
that have a median angle of $\approx31^{\circ}$. As a median value for all
halos we find $\approx24^{\circ}$ at redshift $z=0.1$ and a similar value
for $z=1$, namely, $\approx22^{\circ}$, which is not shown here.

The right panel of
Figure \ref{fig:jdmgasstarangle} shows the angle between the angular
momentum vector of the stars within the inner 10\% of the virial radius and the DM within the entire virial radius. The distribution of the
alignment angles looks similar to that of the gas (left panel of Figure \ref{fig:jdmgasstarangle}) for all galaxies. 
The alignment seems poor, with median angles of $\approx55^{\circ}$ for all halos, $\approx46^{\circ}$ for disks, and
$\approx57^{\circ}$ for spheroidal galaxies. 
This agrees well with \citet{Hahn10}, who calculated a median angle of the stellar
component and the total halo content of about $\approx49^{\circ}$ at $z=0$
for disk galaxies, similar to \citet{Croft09} reporting a median angle of
$\approx44^{\circ}$ at redshift $z=1$. \citet{Bett10} found slightly smaller angles of
$\approx34^{\circ}$ for the alignment of the galaxies with respect to the
hosting total DM halo. 
The galaxies at $z=1.2$ investigated by \citet{Codis15}, using the Horizon-AGN simulation, are slightly less aligned with their DM halos than our simulated galaxies (see also the right panel of Figure \ref{fig:mfrac+alignment} of Appendix \label{sec:appclass}). 
\citet{Deason11} reported that 41\% of their disk galaxies had misalignment angles larger than
$45^{\circ}$ with their DM halo. 
In addition, they found that the disk galaxies are better aligned with the DM in the
innermost 10\% of the $r_{200}$. 
Though we do not show this, here we obtained a median angle of $\approx9^{\circ}$ for the angular
momentum vectors of the stellar and DM components in the
innermost 10\% for our disk galaxies at $z=0.1$, and thus we
can confirm this trend. 
It also agrees well with \citet{Hahn10}, who find a median value of $\approx 15^{\circ}$ for the angle between the DM and stellar components of the galactic disks. For the DM and the gaseous components of their disks they obtain a median value of $\approx 18^{\circ}$, which is in line with the median value of $\approx 12^{\circ}$ for our disk galaxies.


\section{The Host Halos of Different Galaxy Types}

So far we have seen the intrinsic difference of the baryonic components
and how they reflect global halo properties for galaxies classified as
either disk or spheroidal galaxies. We now want to study whether there are
underlying differences in the DM halo contributing to the formation
of these two classes of galaxies.

\subsection{The Alignment of the Host Halos with Their Centers}

\begin{figure}
 \begin{centering}
  \includegraphics[trim={-1cm 3.5cm 1cm 5.2cm}, width=0.45\textwidth, clip=true]{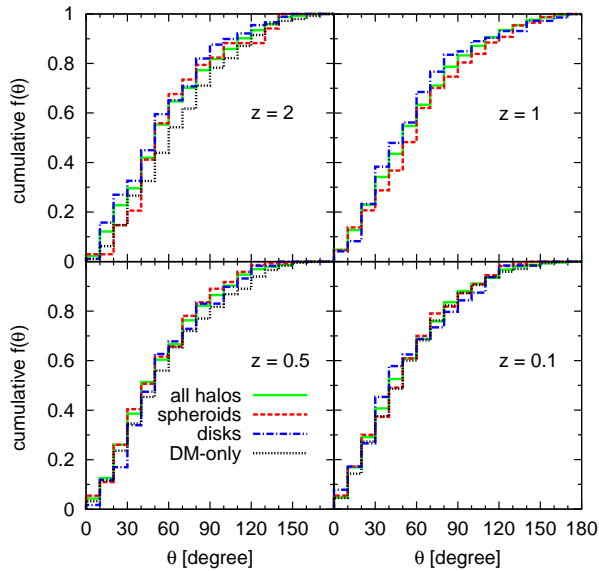}
  \caption{Angle between the total angular momentum vector of
   the DM in $R_\mathrm{vir}$ and that of the DM
   within the innermost 10\% of the virial radius. At
   redshifts $z=2$, $z=0.5$, and $z=0.1$ we overplot the
   distribution of the DM-only run (black dotted). }
  \label{fig:jdmangle_1}
 \end{centering}
\end{figure}

One interesting question is if there are signatures of differences in the internal structure of the angular momentum in the DM component for disk galaxies and spheroidal galaxies. Therefore, we compare the angular momentum of the DM within the whole halo with the one within 10\% of the virial radius.
Figure \ref{fig:jdmangle_1} shows the cumulative distributions of the angle between the DM angular momentum vector of the whole halo and that of the inner part of the halo.
At $z=2$, $z=0.5$, and $z=0.1$ we include the corresponding distribution of the
DM-only run \footnote{For technical reasons there were no data available at $z=1$ from the DM-only run.}. 
The alignment is generally better for disks than for spheroids (see also Table \ref{tab:angles_dm_dm10}), as already speculated in \citet{B01}. 
Interestingly, the relative difference between the alignments for disk and spheroidal galaxies is largest at $z=1$, 
which corresponds to a typical formation redshift of spiral galaxies.
Overall,
the misalignment grows with redshift for all halos. At redshift $z=0.1$ we calculate median values for all
halos of $\approx47^{\circ}$, which agrees well with \citet{Hahn10}, reporting a value of $45^{\circ}$.

\begin{table}
 \caption{Median Misalignment Angles of the DM at Different Radii}
 \centering
 \label{tab:angles_dm_dm10}
 \begin{tabular}{l c c c c }
 \hline \hline
  Redshift & All Halos & Disks & Spheroids & DMO \\ \hline
  2 & 56.7 & 54.8 & 57.2 & 66.6\\ 
  1 & 54.9 & 51.2 & 61.3 & --- \\
  0.5 & 48.3 & 53.1 & 48.8 & 53.5\\
  0.1 & 47.4 & 42.2 & 50.6 & 51.3\\
 \hline  
 \end{tabular}
\tablecomments{The median misalignment angles between the angular momentum vectors of the DM within 10\% and that of the total DM halo ($\theta_\mathrm{DM-DM10}$) for the hydrodynamical run as well as for the run with only DM (DMO).} 
\end{table}

In contrast, in the DM-only run the values are slightly higher at all redshifts, with
median misalignment angle of $\approx51^{\circ}$ at $z=0.1$.
This tendency was also seen by
\citet{Bett10}, who found that in their run with baryons the vectors
were slightly more aligned than in the DM-only case. 
The misalignment angle in our analysis is higher than their median angles of
$15^{\circ} - 30^{\circ}$ for their run with baryons. This could be
due to the fact that we only consider the inner 10\% instead
of 25\%, as in their study. 
This is expected, since \citet{BS05} found that
the alignment becomes worse when the radii are further separated.

A possible interpretation of the above findings could be that disk galaxies
preferentially reside in halos, where the core is better aligned with
the outer parts of the halo \citep[see also][]{B01}. 
It might well be that in such halos the angular momentum can be 
transported more effectively by the cooling of gas from the outer
parts into the central parts. Another possibility could be that
disk galaxies survive merger events longer, when consecutive
infall is aligned with the angular momentum of the galaxy.
Interestingly, spheroidal galaxies (especially at $z\approx1$) show
exactly the opposite behavior. The inner and outer
parts of their DM halos are less aligned,
indicating that major merging events are contributing 
to the buildup of spheroidal galaxies. 
This is in line with previous studies showing that major
mergers with misaligned spins can be responsible for angular
momentum misalignments \citep{SSBH12}.

In addition, \citet{Welker14} proposed that anisotropic cold streams realign the galaxy with its hosting
filament. However, \citet{Sales12} suggested that disks form out of
gas having similar angular momentum directions, which would favor the
spherical hot accretion mode, while the accretion along cold flows
that are mainly misaligned tends to build a spheroid. To answer this
question in detail, further investigation is needed to trace back
disk galaxies and see whether the primordial alignment causes the inflowing
matter to become a spiral or whether it is the other way around, i.e., if the
galaxy type over cosmic time induces the alignment. We suspect that the
environment has a significant impact on the formation of disk galaxies.
In less dense environments, where the halos can evolve relatively
undisturbed, the angular momentum of the galaxy preferably remains aligned with
that of its host halo, and thus a disk can form. 


\subsection{The Spin Parameter $\lambda$ of the Host Halos}\label{subsec:lambda}

\begin{figure}
\centering
\includegraphics[trim={0cm 3.5cm 1cm 5.2cm}, width=0.45\textwidth, clip=true]{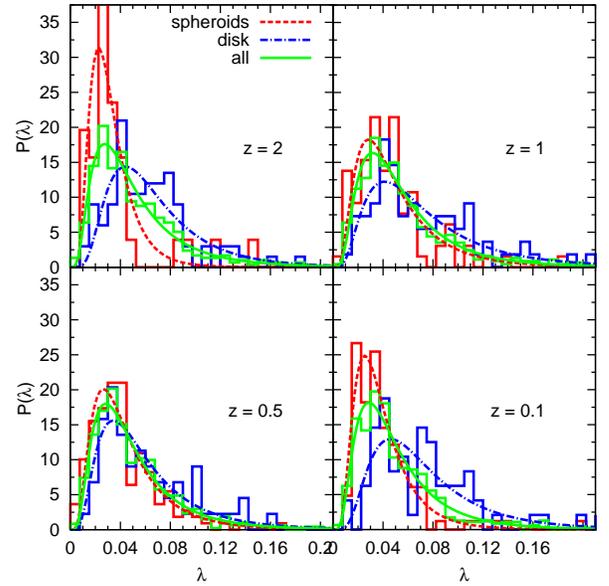}
\caption{The $\lambda$-distribution calculated with formula
 \eqref{eq:lambda} for the different redshifts as indicated in the
 plots. Green takes all halos into account. The distributions of the
 spheroids are the red dashed histograms, and those of the disks are
 blue dot-dashed histograms. The curves are the fits given by equation
 \eqref{eq:lambda_fit}. At all redshifts a split-up of the two
 different galaxy types is present. The spheroids have their median
 at the lower $\lambda$-values, while the disks have higher values. }
\label{fig:lambda}
\end{figure}

In the following section we finally return to the spin parameter $\lambda$,
evaluated for the whole halo. 
In Figure \ref{fig:lambda} we show the distribution of the $\lambda$-parameter
for the total matter distribution within our halos for different redshifts. 
The histograms and lognormal fit curves (see Equation \ref{eq:lambda_fit}) are each normalized to the number of the halos. 
The distribution of $\lambda$ for all 396 halos at $z=2$ (upper left), 
606 halos at $z=1$ (upper right), 629 halos at $z=0.5$ (lower left), and 621 
halos at $z=0.1$ (lower right) is shown in green. 
We also plot the distribution of the spheroidal (red dashed) and disk (blue dot-dashed) galaxies. 
The spheroids, which have only little rotation
in the stellar component, tend to have lower $\lambda$-values. 
The disk galaxies have
their median at higher values. 
This bimodality of the two galaxy types is seen at all four redshifts, as already shown in \citet{Teklu15}. The spheroids have always lower median values than the disk galaxies (see also Table \ref{tab:lambda} for the fitting parameters). 
Observationally, \citet{Hernandez07} also found that the spiral galaxies on average have larger $\lambda$-values than the ellipticals of their sample. 
However, this was not seen in previous studies by \citet{Sales12} and \citet{Scan09}, who did not find a correlation between
galaxy type and the spin parameter. 

\begin{table}
\caption{Calculated Median Values and the Fit Values of the $\lambda$-Distributions}
 \label{tab:lambda}
\centering
\begin{tabular}{l l}
\hline \hline
  Redshift & All Halos \\ \hline
  2 & $\lambda_\mathrm{med}=0.043$, $\lambda_{0}=0.043$, $\sigma=0.662$ \\
  1 & $\lambda_\mathrm{med}=0.046$, $\lambda_{0}=0.047$, $\sigma=0.635$ \\
  0.5 & $\lambda_\mathrm{med}=0.041$, $\lambda_{0}=0.042$, $\sigma=0.640$\\
  0.1 & $\lambda_\mathrm{med}=0.042$, $\lambda_{0}=0.043$, $\sigma=0.630$ \\ \hline
  Redshift & Disks \\ \hline
  2 & $\lambda_\mathrm{med}=0.058$, $\lambda_{0}=0.059$, $\sigma=0.543$\\
  1 & $\lambda_\mathrm{med}=0.060$, $\lambda_{0}=0.062$, $\sigma=0.644$ \\
  0.5 & $\lambda_\mathrm{med}=0.051$, $\lambda_{0}=0.050$, $\sigma=0.617$\\
  0.1 & $\lambda_\mathrm{med}=0.069$, $\lambda_{0}=0.064$, $\sigma=0.570$\\ \hline
  Redshift & Spheroids \\ \hline
  2 & $\lambda_\mathrm{med}=0.028$, $\lambda_{0}=0.029$, $\sigma=0.497$\\
  1 & $\lambda_\mathrm{med}=0.042$, $\lambda_{0}=0.042$, $\sigma=0.633$\\
  0.5 & $\lambda_\mathrm{med}=0.037$, $\lambda_{0}=0.039$, $\sigma=0.620$\\
  0.1 & $\lambda_\mathrm{med}=0.034$, $\lambda_{0}=0.034$, $\sigma=0.546$\\
\hline  
\end{tabular}
\tablecomments{The calculated median values $\lambda_\mathrm{med}$ and the
 fit values $\lambda_{0}$ and $\sigma$ of the $\lambda$-distributions
 at different redshifts for our sample of galaxies.}
\end{table}
The $\lambda$-distribution for all halos seems to be relatively
constant with time, i.e., independent of redshift, which is in
agreement with \citet{Peirani04}. 

\begin{figure}
\centering
\includegraphics[trim={0cm 3.5cm 1cm 5.2cm}, width=0.45\textwidth,clip=true]{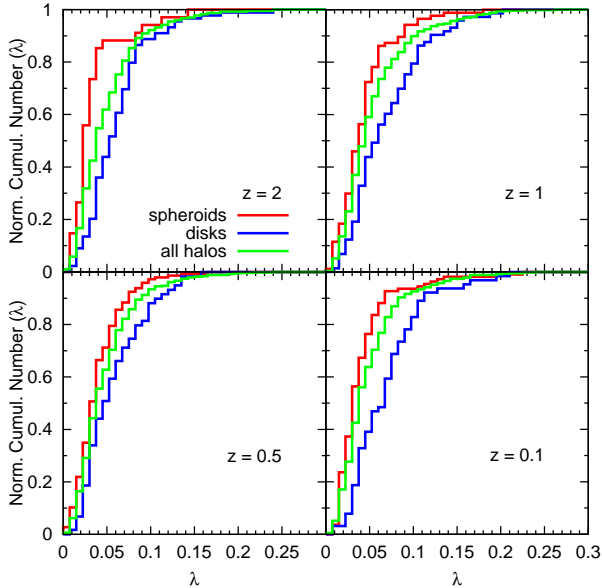}
\caption{The $\lambda$-parameter plotted against the normalized
 cumulative number of halos. Green takes all halos into account,
 red the spheroidal galaxies, and blue the disk galaxies. The distributions of
 the disks and the spheroids do not overlap (besides at $z=0.1$ at
 the higher end). The disks are on the right side, i.e., at higher values,
 of the distribution for all halos, while the spheroids are on the left,
 i.e., at lower values. }
\label{fig:lambda_cum}
\end{figure}

To verify the statistical significance of the differences, 
we show in Figure \ref{fig:lambda_cum} the cumulative distributions
for the $\lambda$-values, again splitting the halos according to the different
galaxy types they host. 
This illustrates the distances between the two distributions, where the curve for the halos of the spheroidal galaxies is always left of the curve for all halos and the curve of the disk galaxies stays always to the right.
The differences are more pronounced for small spins.
To quantify the statistical significance of differences in the distributions for the disk and spheroidal galaxies, we apply a Kolmogorov-Smirnov (K-S) test. 
This test is applied to the two unbinned distributions for the disks and the
spheroids. Table \ref{tab:ks} shows the calculated values for
the maximum distance $D$ and the probability. With the exception of the
results\footnote{Although unlikely, the value obtained at $z=0.5$ does not allow to exclude the same origin
of the two distributions.} at $z=0.5$, this confirms that the spin distributions 
of the halos hosting disk galaxies are statistically significantly different from
the halos hosting spheroidal galaxies. 

\begin{table}
\caption{K-S Test for the $\lambda$-Distributions}
 \label{tab:ks}
 \centering
\begin{tabular}{l l l}
\hline \hline
  Redshift & $D$ & Probability \\ \hline
  2 & 0.563 & 1.41 $\cdot 10^{-7}$ \\
  1 & 0.319 & 4.45 $\cdot 10^{-4}$\\
  0.5 & 0.237 & 1.45 $\cdot 10^{-2}$\\
  0.1 & 0.421 & 6.31 $\cdot 10^{-7}$ \\
\hline  
\end{tabular}
\tablecomments{The maximum distance $D$ and the significance level
 (probability) resulting from the K-S test that the
 $\lambda$-distributions for disks and spheroids at different
 redshifts for our sample of galaxies are from the same distribution.}
\end{table}

In this section we have seen that there is a statistical correlation between the morphological type and the overall distribution of the spin parameter $\lambda$. 
We thus conclude that the total DM halo somehow ``knows'' about the morphology of the galaxy at its center. 
At all four considered redshifts the distributions for the spheroids
have lower median $\lambda$-values than those of the disks. According to the K-S test, there is a strong indication that they do not originate from the
same distribution. On the other hand, we have also found that there
are spheroids with even higher $\lambda$-values than disk galaxies. 


\begin{figure}
\centering
\includegraphics[trim={0cm 3.5cm 1cm 5.2cm}, width=0.45\textwidth]{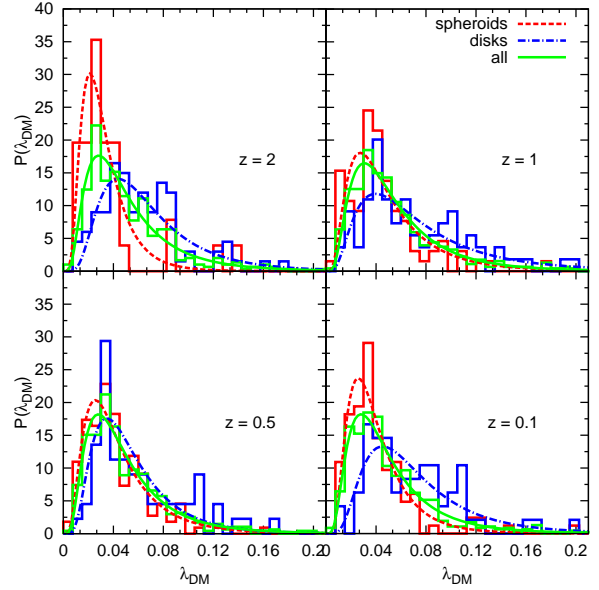}
\caption{The $\lambda$-parameter for the DM component within
 $R_\mathrm{vir}$; also in the DM component of all halos
 (green) there is a broad distribution, which splits into spheroids
 (red) at lower values, while disks (blue) peak at higher values.}
\label{fig:lambda_dm}
\end{figure}

To verify that these differences are intrinsic to the halo and not
caused by the contributions of the baryonic component to the spin
parameter, we calculated the spin parameter $\lambda_\mathrm{DM}$ for
the DM component of the halos. Figure \ref{fig:lambda_dm}
shows the $\lambda_\mathrm{DM}$-distribution for the baryon run for
the four redshifts as indicated in the plots. The green histogram
shows $\lambda_\mathrm{DM}$ for all halos, the red one stands for the
spheroids, and the blue one for disk galaxies. Here again, most prominently visible
at redshifts $z=2$ and $z=0.1$, there is a splitting of the two
different galaxy types, the distribution of the spheroids peaks at
lower values, and that of the disks at higher values. We also perform a
K-S test on the two distributions (see Table \ref{tab:ks_dm}). The
values are similar to that of the general $\lambda$-distribution
(Table \ref{tab:ks}). 

\begin{table}
\caption{K-S Test for the $\lambda_\mathrm{DM}$-Distributions}
 \label{tab:ks_dm}
 \centering
\begin{tabular}{l l l l l}
\hline \hline
  Redshift & $D$ & Probability & $D_\mathrm{DMO}$ & $\mathrm{Probability}_\mathrm{DMO}$  \\ \hline
  2 & 0.537 & 5.92 $\cdot 10^{-7}$ & 0.590 & 8.14 $\cdot 10^{-8}$ \\
  1 & 0.319 & 4.45 $\cdot 10^{-4}$ & --- & --- \\
  0.5 & 0.234 & 1.65 $\cdot 10^{-2}$ & 0.285 & 2.3 $\cdot 10^{-3}$\\
  0.1 & 0.390 & 4.60 $\cdot 10^{-6}$ & 0.370 & 9.84 $\cdot 10^{-5}$\\
\hline  
\end{tabular}
\tablecomments{The maximum distance $D$ and the significance level
 (probability) resulting from the K-S test that the
 $\lambda_\mathrm{DM}$-distributions for disks and spheroids at
 different redshifts originate from the same distribution for the baryon run and the DM-only (DMO) run.}
\end{table}

In the Appendix \ref{sec:applam} we show that this split-up of the galaxy types is also reflected in the spin parameter $\lambda$ of the stellar component.


In order to see whether the differences originate from differences within the DM halo or if they are caused by
the interplay between the DM and the baryonic component, 
we finally compare the run with the baryons to the DM-only control run. 
We thereto cross-identify the halos of the DM-only run with those of the hydrodynamical run. 
We search for the corresponding halos in the DM run, allowing the center
of the halo to be in a range of 200 kpc around its position in the simulation with baryons, and additionally restrict to halo pairs that only 
differ by up to 30\% in their virial masses. 
At redshift $z=2$ we found 364 halos in the
baryon run that match with the DM-only run, at $z=0.5$ we could assign 609
halos, and at $z=0.1$ we cross-matched 575 halos.
We find that the $\lambda_\mathrm{DM}$-values of each halo identified in both runs are very similar (see Figure \ref{fig:lambda_dmo_dmb+jrm}  of Appendix \label{sec:applam}). 
Since baryons have an effect on the DM, in the two runs the evolution of some of the halos can be quite different, as also discussed in \citet{Bett10}.
Therefore, we cannot match every halo.

\begin{figure*}
 \begin{centering}
  \includegraphics[trim={4cm -0.5cm 4.2cm 0cm}, width=0.37\textwidth, clip=true, angle=270]{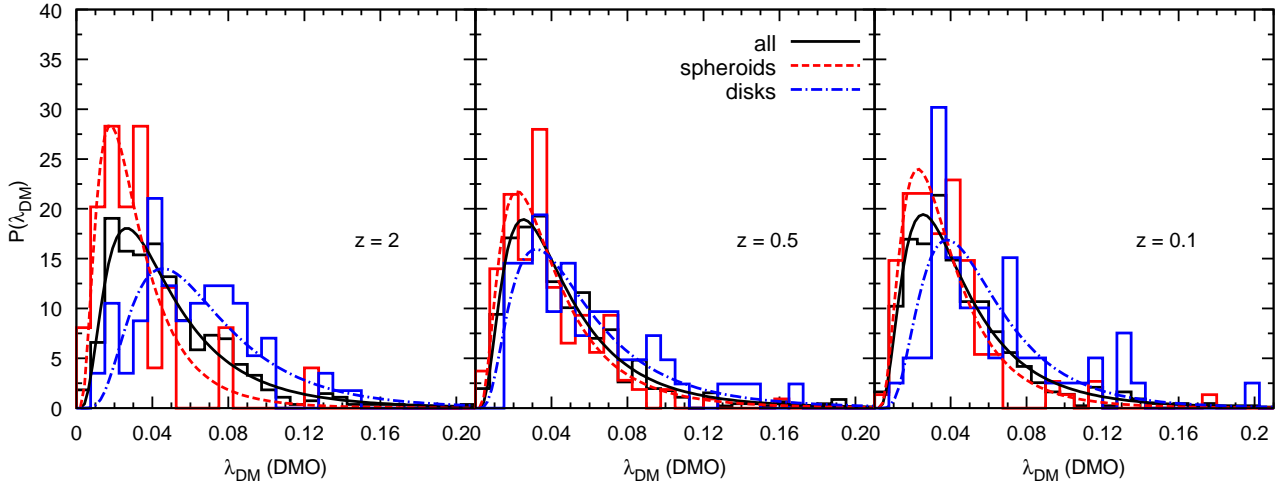}
  \caption{The $\lambda$-parameter for the DM component for
   the DM-only run (black) at three redshifts. The red curves show
   the distribution for halos that were identified in the baryon
   run and classified as spheroids, while the blue curves show
   the halos classified as disks in the baryon run. There is a
   split-up of the galaxy types, which suggests that the morphology
   could be a result of the formation history. }
  \label{fig:lambda_dm_dmo}
 \end{centering}
\end{figure*}

Figure \ref{fig:lambda_dm_dmo} shows the $\lambda$-distribution for the DM component at three different redshifts. 
The overall distribution is shown in black. The fit values $\lambda_{0}$ for the DM-only run are slightly lower than for the run with baryons.
When we split the halos in the DM-only run according to their galaxy
type assigned in the run with baryons, we find a similar split in the distributions
as in the run with baryons. At redshift $z=2$ we could use 76 disks and 33 spheroids,
at $z=0.5$ there are 55 disks and 143 spheroids, and at $z=0.1$ we cross-matched 53 disks and 99 spheroids. 
The results of the K-S test (see Table \ref{tab:ks_dm}) show that these two distributions are unlikely to originate from the same one. 
It is striking that even in the run without baryons, the split-up of the spin parameter for the different galaxy types is clearly visible. 
This suggests that the hosting DM halo and, connected with that, the formation history and environment play an important
role for the morphology of the resulting galaxy. 

\section{Discussion and Conclusions}\label{sec:conclusion}

We extracted between $400$ and $630$ halos with total halo masses above $5\cdot 10^{11} M_{\odot}$ at four different redshifts from the hydrodynamical, cosmological state-of-the-art simulation Magneticum Pathfinder, which includes detailed
treatment of star formation, chemical enrichment, and evolution of
supermassive BHs. We investigated the distribution of the 
spin parameter $\lambda_k$ of the different components (DM, 
stars, hot and cold gas) and the alignment of the angular momentum vectors of those components within the entire halo, as well as within
the central 10\% of the virial radius. 
To classify the galaxies according to their morphology, we rotate them such that their angular momentum vector is oriented along the $z$-axis. 
For this orientation we calculate the circularity parameter $\varepsilon$, which allows a classification based on their circularity distribution. 
This allows us to define the subset of galaxies clearly identified as spheroidal or disk galaxies and to compare their properties to observations. 
We additionally performed and analyzed a DM control simulation to test the effect of the baryonic processes and the formation history on the angular
momentum within the halos. We summarize our findings as follows:

(i) For all our halos, the stellar component generally has a lower
  spin than the DM component, while the gas shows a
  significantly higher spin parameter, especially the cold gas component that dominates the overall spin parameter of the gas. While the
  distribution of the spin parameters of stars and DM does not show a significant evolution with time, the spin
  distribution of the gas component significantly evolves toward
  larger spin values with decreasing redshift. 
  
(ii) In general, the angular momentum vectors of the baryonic components and the DM are well aligned. There is an evolution within these
  alignments, where at high redshifts the alignment between gas and
  stars is better than their alignment with the DM component,
  while at low redshift stars and DM tend to be better aligned. 
  
(iii) When classifying galaxies according to their position within the stellar-mass--specific-angular-momentum plane, we demonstrate that various galaxy properties show smooth transitions, as expected when going from rotation-dominated systems to dispersion-dominated systems. This is most prominently found for the
  circularity distribution of stars and gas, the cold gas fraction, but
  also for the spin parameter distribution and the alignments of the
  angular momenta. 
  Rotation-dominated galaxies have generally larger spin parameters, and the angular momentum vectors of the stars and gas components are aligned. 
  On the contrary, dispersion-dominated galaxies generally have smaller spin parameters and show almost no alignment between the angular momentum vectors of stars and the (however small) gas components. 
  
(iv) Alternatively, when classifying our galaxies according to their circularity
  distribution in combination with the cold gas content into disk
  galaxies and spheroidal galaxies, they populate clearly distinguishable regions within the stellar-mass--specific-angular-momentum plane. 
  Using that classification, we lose a significant number of galaxies that cannot be classified within this scheme. Nevertheless, for those galaxies that can be classified the results are in excellent agreement with observations for both spheroidal and disk galaxies. 
  
(v) In disk galaxies the specific angular
  momentum of the gas is slightly higher than that of the stars, which is
  in good agreement with recent observations \citep{Obreschkow14}.
  This is due to the fact that the gas component has contributions
  from freshly accreted gas with higher specific angular momentum,
  which will be turned into stars later. This is also reflected in our
  result that the specific angular momentum of the young stars is slightly larger compared to
  that of all stars.
  
(vi) In general, the specific angular momentum of the total halo is
  higher than the specific angular momentum of the gas in the galaxy.
  For disk galaxies it accounts for roughly 43\% of the value found 
  for the total halo, with no significant redshift evolution.
  
(vii) Overall, the angular momentum of the total halo is only
  weakly aligned with the angular momentum of the central part. Here
  the simulation including baryons shows slightly more alignment than
  the DM-only simulation. 
  However, we found that in general the halos hosting disk galaxies (in comparison with
  halos with spheroidal galaxies at their centers) show a better alignment of the angular momentum vectors of the total halo and the central part for the DM component. 
  This is most pronounced at redshift $z\approx1$, where most disk galaxies are forming.
  
(viii) The splitting of the galaxies into disk and spheroidal galaxies
  reveals also a dichotomy in general halo properties: the halos
  hosting disk galaxies have a slightly larger spin than the halos
  hosting galaxies classified as spheroidal galaxies. This dichotomy
  is even reflected in the distribution of the spin parameters in the
  DM control run, where we cross-identified the halos which
  in the hydrodynamical simulation host galaxies of different
  types. This indicates that the formation history of the DM
  halo plays an important role for defining the morphology of the
  galaxies.

Our results are based on the classification of galaxies obtained from
the circularity of the stellar component ($\varepsilon_\mathrm{star}$)
in combination with the fraction of cold gas with respect to the
stellar mass within the central part of the halo, where the galaxies
form. This allows us to select classical disk and spheroidal
galaxies, which show very distinct dynamical properties, reflected in
their circularity distribution, cold gas fractions, specific angular
momentum, and spin and angular momentum vector alignments. They therefore
shed light on the main formation mechanism of these galaxies within
the cosmological framework. 

While our current classification focuses on a selection of galaxies whose properties
resemble those of classical spiral and elliptical galaxies, the lion's share of the
galaxies in our simulation cannot be classified as poster child disk or spheroidal
galaxies.
This again reflects observational facts, since at present day many galaxies exhibit
signs of distortions, ongoing merger events, peculiar structures, or other
irregularities.
For our unclassified galaxies, we clearly see that their properties show a smooth
transition between the poster child disks and spheroids, and to understand those
transition processes, more complex classification schemes are needed.
We suggest that those schemes should consider a differentiation between young and
old stellar components, as well as the general distribution of the diverse gas phases
in different parts of the galaxies.

Additionally, when comparing to observations, more emphasis needs to
be placed on how observational quantities are obtained and how this is
mimicked when analyzing simulations. Classifications of galaxy morphologies
in cosmological simulations also allow the interpretation of cosmological
simulations in various new ways, as already proven in this work, not only for studying the formation
and evolution of galaxies but also for studying the relation of AGNs
with their host galaxies, as well as for cosmological studies, where
the bias of the measurements depends on the morphologies of the galaxies.
Therefore, the good agreement of the intrinsic dynamical properties
for our classified galaxies with respect to current observations can
be seen as a first step to promote cosmological, hydrodynamical
simulations for future cosmological studies.


\acknowledgments
We thank the anonymous referee for helpful comments. 
AFT, KD and AMB are supported by the DFG Research Unit 1254.
AB is supported by the DFG Priority Programme 1573.
AFT, KD and LKS are supported by the DFG Transregio TR33. 
This research is supported by the DFG Cluster of Excellence ``Origin and Structure of the Universe.'' 
We are especially grateful for the support by M. Petkova through the Computational Center for Particle and Astrophysics (C2PAP). 
Computations have been performed at the `Leibniz-Rechenzentrum' with CPU time assigned to the Project ``pr86re.''
We thank Michael Fall for helpful comments. 


\appendix

\section{Appendix A: Details on the Classification}\label{sec:appclass}

\begin{figure*}
 \begin{centering}
 \centerline{\includegraphics[trim={-0.5cm 3.5cm 1cm 5.2cm}, clip=true, width=0.45\textwidth]{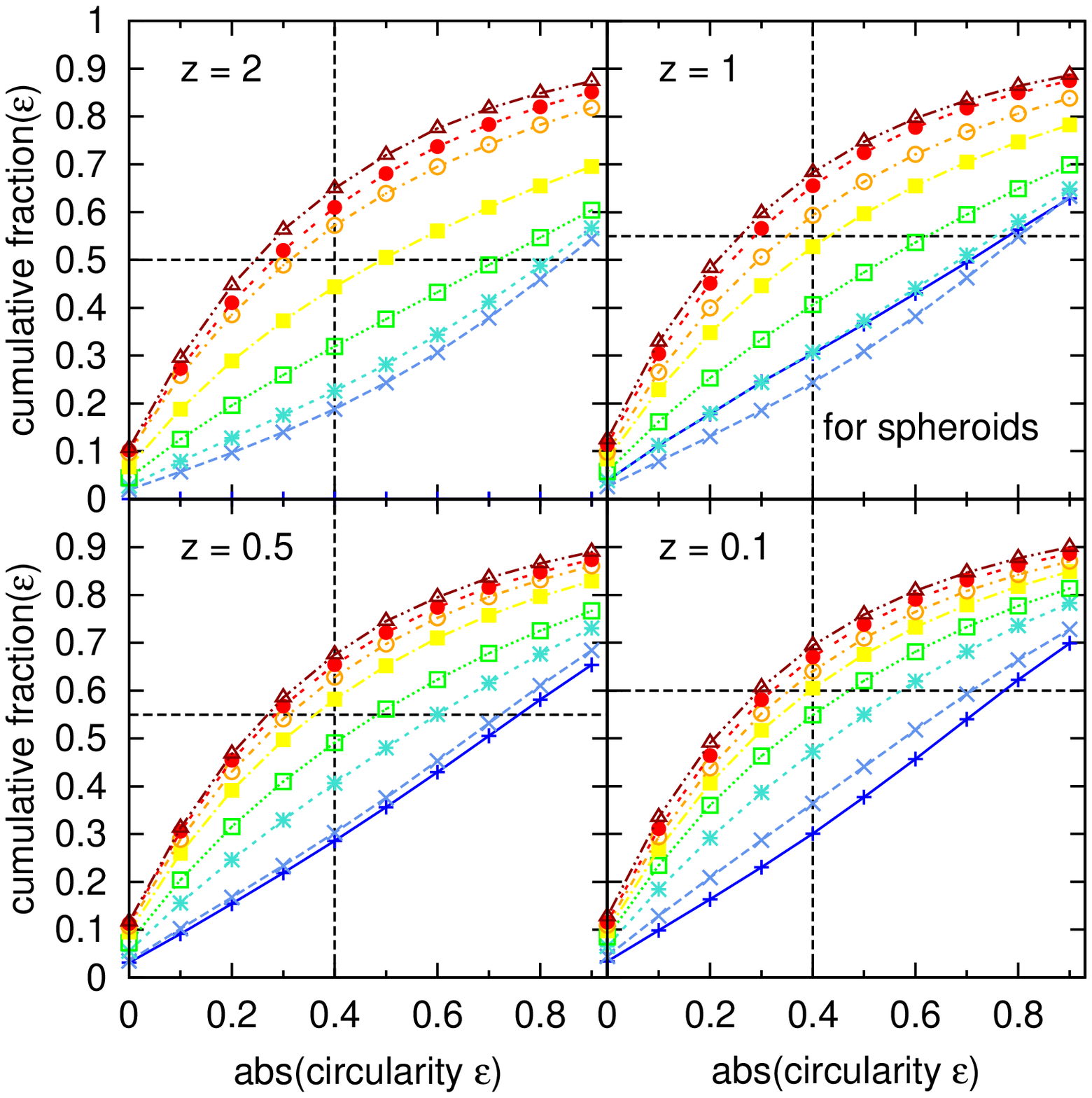}
             \includegraphics[trim={-0.5cm 3.5cm 1cm 5.2cm}, clip=true, width=0.45\textwidth]{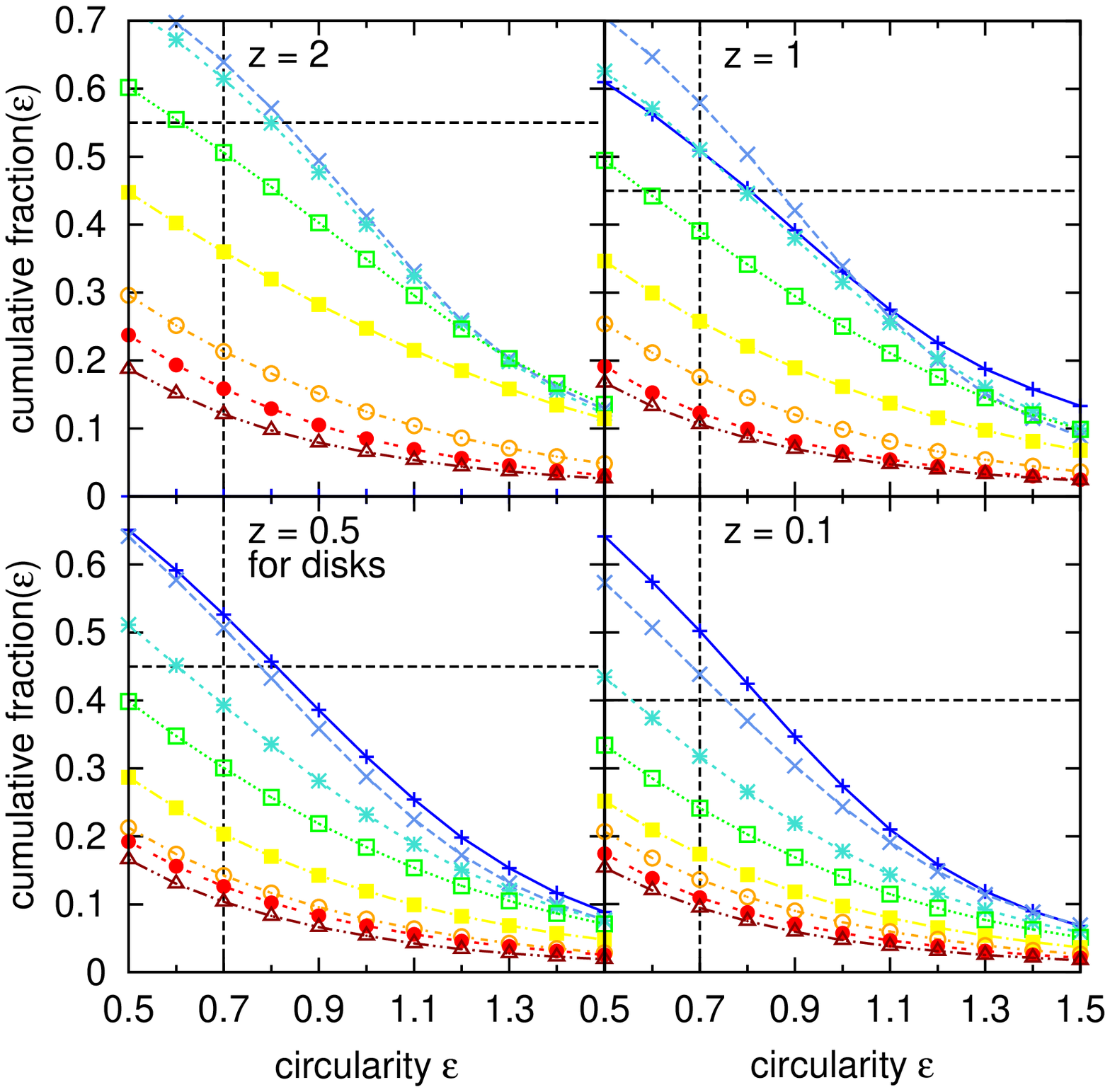}}
 \caption{Left: cumulative fraction of all galaxies in dependency of the circularity $\varepsilon$ of the stars for all four redshifts. 
 The colors reflect the different $b$-value bins, as in Figure \ref{fig:av_eps_bin}. 
 From this we estimate the cut for the determination of the spheroidal part of a galaxy. This cut differs for each redshift. Right: cumulative fraction in order to determine the disk part of a galaxy. The values on the $y$-axis are the cumulative values from the corresponding point on the $x$-axis up to an $\varepsilon$-value of 3. }
 \label{fig:cum_eps}
 \end{centering}
\end{figure*}

In order to illustrate the choice of the cuts in $\varepsilon_\mathrm{star}$ used for the classification of our galaxies, we show the cumulative circularity distribution for spheroids (left panels) and the `anti-'cumulative distribution for disk galaxies (right panels) in Figure \ref{fig:cum_eps}. 
For the spheroids we sum up the fractions from $-\varepsilon$ to $\varepsilon$ of the distributions shown in Figure \ref{fig:av_eps_bin}, $f(x)=\sum_{-x}^{x} fraction(\varepsilon)$. The cut is drawn between the halos of the $b$-value-bin that show a clear behavior of spheroids at the corresponding redshift. 
The same is done for the disk galaxies, besides that here we sum up the fraction from $\varepsilon$ to $3$, $f(x)=\sum_{x}^{3} fraction(\varepsilon)$.

\begin{figure}
\centerline{\includegraphics[height=7cm, clip=true]{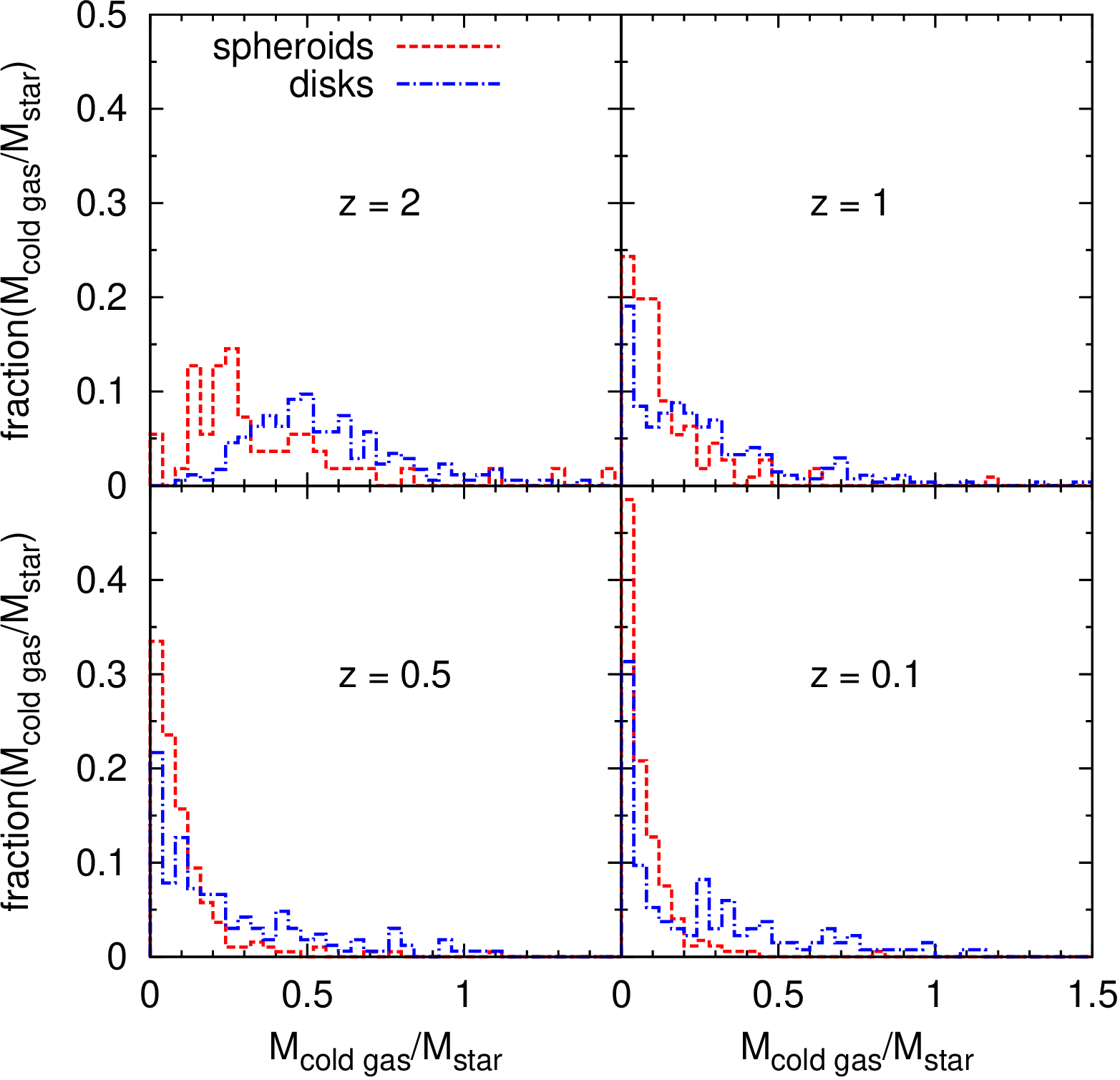}\hspace{1cm}
 \includegraphics[height=7cm, clip=true]{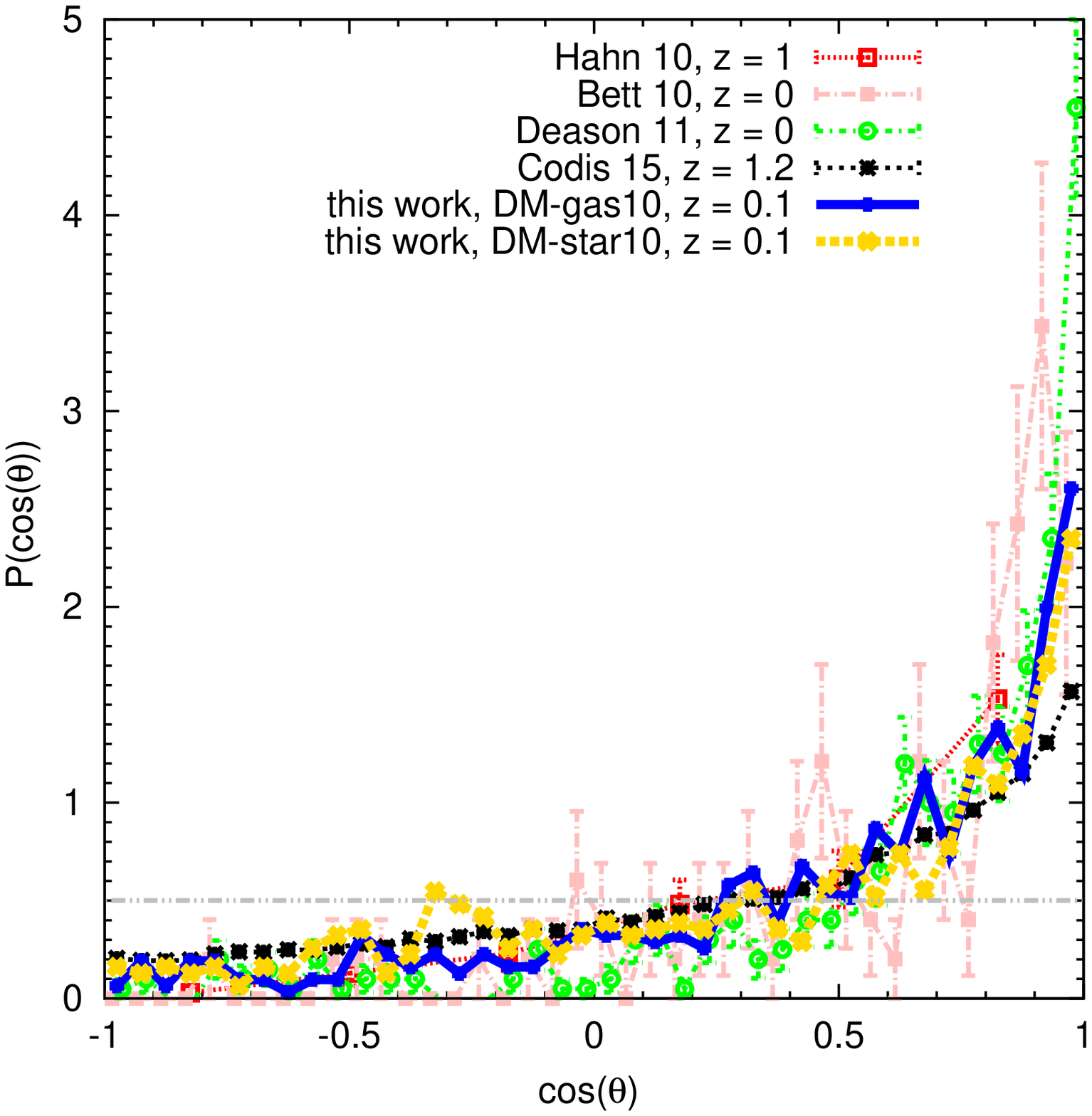}}
\caption{Left: fraction of the mass of the cold gas with respect to the stellar mass, both within the inner 10\% of the virial radius. 
The cold gas fraction decreases continuously with decreasing redshift, for spheroids faster than for disk galaxies. 
Note that at $z=2$ most spheroidal galaxies have a mass fraction larger than 10\%. 
At $z=0.1$ the disk galaxies are divided into two populations, one with a fraction higher than 30\% and the other having less than 10\%. 
We suggest that the second population comprises lenticular galaxies.
Right: probability distribution function of the alignment of the galaxies compared to their DM halo in comparison with previous studies, as indicated in the plot.  
}
  \label{fig:mfrac+alignment}
\end{figure}

In the left panels of Figure \ref{fig:mfrac+alignment} we show that at high redshift there are many spheroidal galaxies (red histograms, classified with $\varepsilon_\mathrm{star}$) with a high fraction of cold gas mass with respect to the stars. 
This fraction decreases very fast with decreasing redshift. 
Disk galaxies (blue) have a higher amount of cold gas at all times, especially at high redshift. 
Still, at low redshift there are many disk galaxies that have less than 10\% cold gas. 
Those galaxies actually resemble S0 properties and are not classical ``disks.''

\begin{table}
\caption{Upper/Lower Limits for the Classification Criteria}
\label{tab:class_cut}
\centering
\begin{tabular}{l l l l l}
\hline \hline
  Redshift & $\varepsilon_\mathrm{star}$, $d$ & $\varepsilon_\mathrm{star}$, $s$ & $M_\mathrm{cold}/M_\mathrm{star}$, $d$ & $M_\mathrm{cold}/M_\mathrm{star}$, $s$ \\ \hline
  2 & 0.55 & 0.5 & 0.5 & 0.35 \\
  1 & 0.45 & 0.55 & 0.35 & 0.2 \\
  0.5 & 0.45 & 0.55 & 0.275 & 0.125\\
  0.1 & 0.4 & 0.6 & 0.215 & 0.065\\
\hline  
\end{tabular}
\tablecomments{The values for the classification depending on redshift. We list the lower limit for the $\varepsilon_\mathrm{star}$ for disks ($d$) as well as spheroids ($s$). For the disks the $M_\mathrm{cold}/M_\mathrm{star}$ is a lower limit, while the spheroids have to have less than the given value.}
\end{table}

In Table \ref{tab:class_cut} we list the values of the cuts for the classification criteria. The values for the spheroids should be understood as an upper limit, while for the disks they represent a lower limit.


\section{Appendix B: The Alignment of the Galaxy with Its DM Halo}\label{sec:appalign}

The right panel of Figure \ref{fig:mfrac+alignment} shows the probability distribution function of the cosine of the angles between the innermost 10\% of the virial radius of our disk galaxies and the hosting DM halo. The solid blue line stands for the gaseous component, and the yellow dashed line for the stellar component. 
We extracted the other data points from \citet{Codis15} for an easier comparison with previous works. 
Overall, our results are in good agreement with previous studies. We find that our disk galaxies have a slightly weaker alignment than those of \citet{Bett10} and \citet{Deason11}. Our results agree well with \citet{Hahn10}. \citet{Codis15} report slightly less aligned galaxies.


\section{Appendix C: The Spin Parameter $\lambda$}\label{sec:applam}

\begin{figure}
 \centerline{\includegraphics[height=7cm, clip=true]{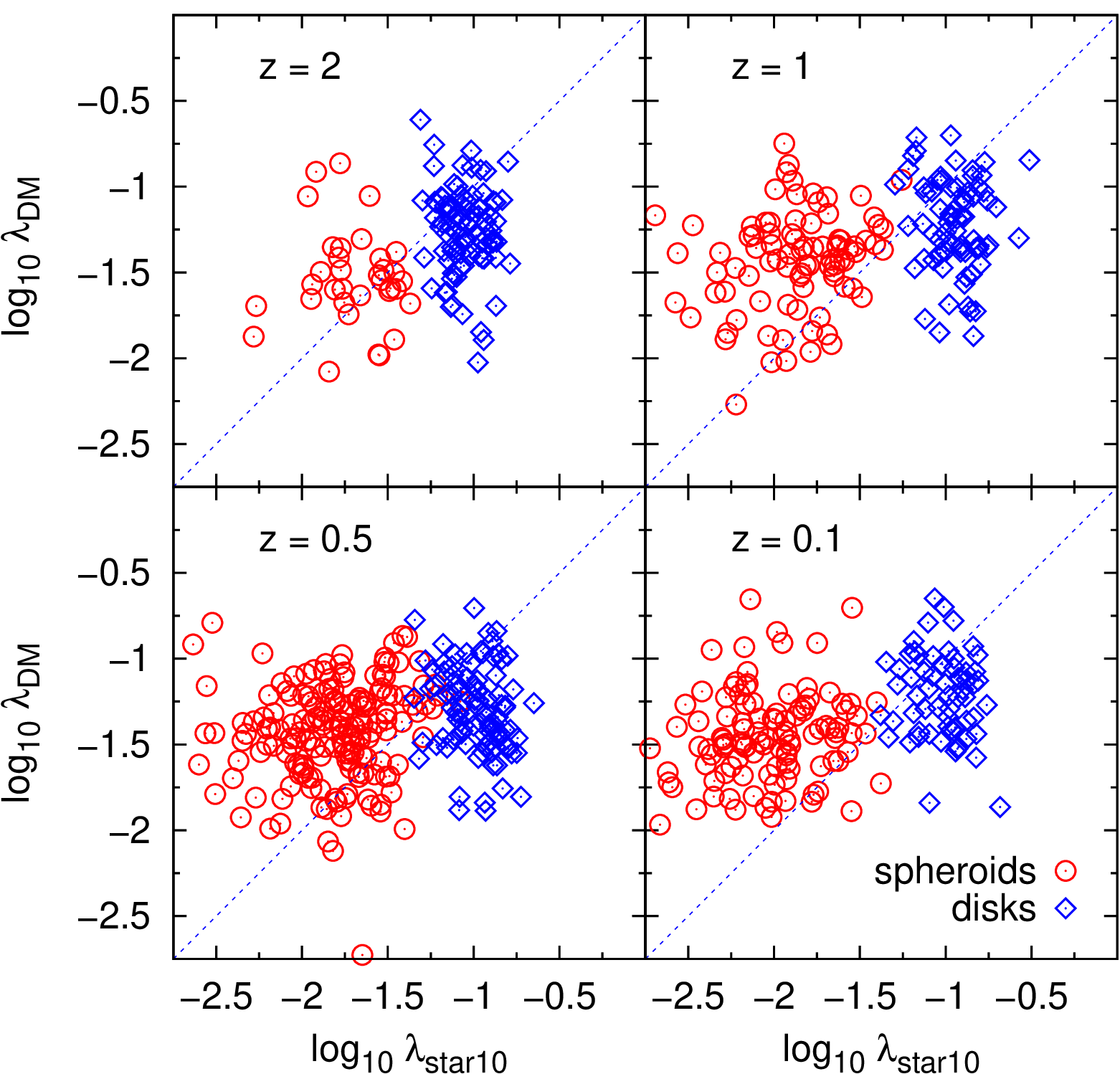}\hspace{1cm}
 \includegraphics[height=7cm, clip=true]{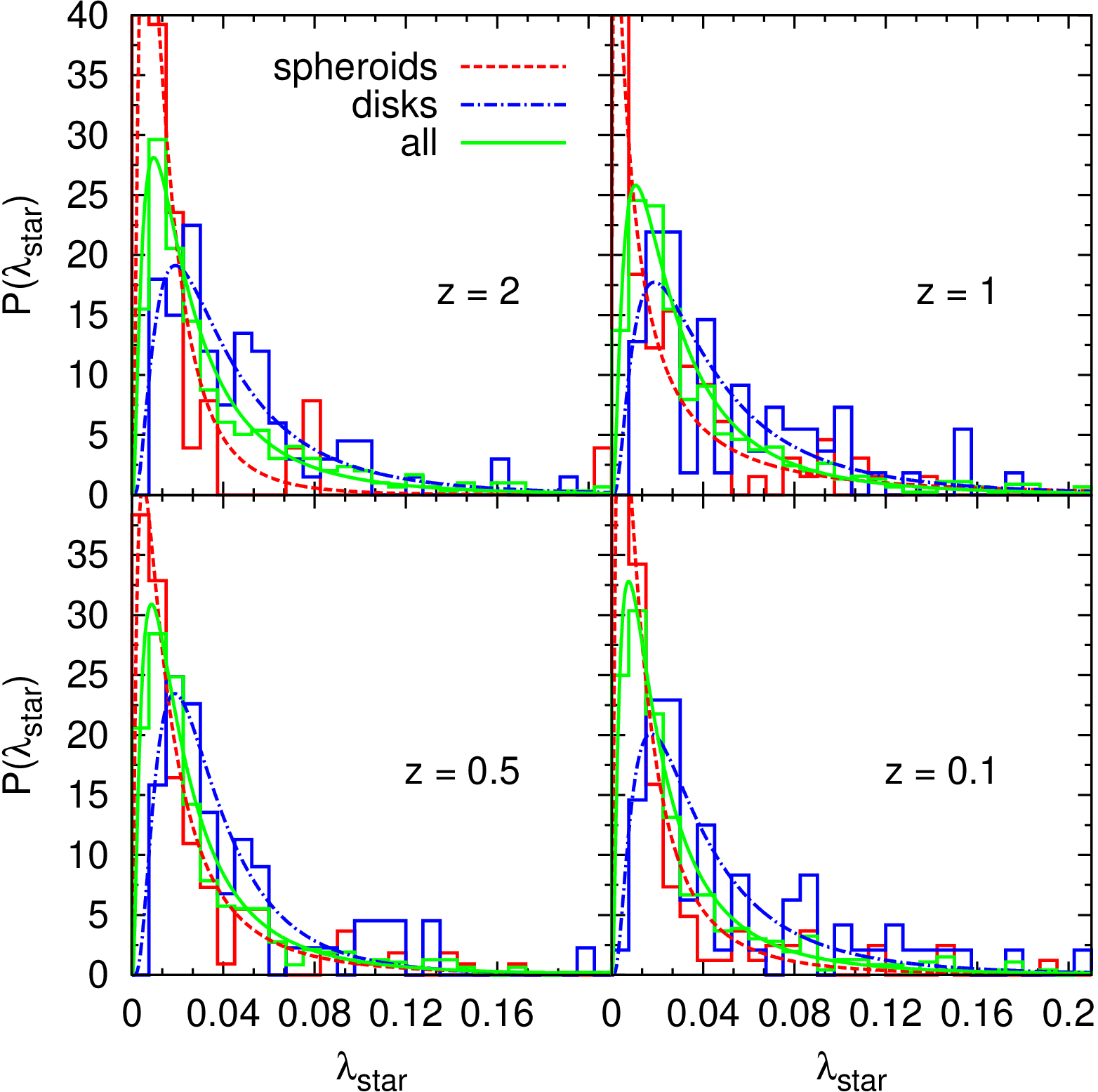}}
 \caption{Left: $\lambda$-parameter for the stellar component in the innermost 10\% against that of the DM of the entire DM halo, at four different redshifts, divided into spheroidal (red circles) and disk (blue diamonds) galaxies. 
 Right: $\lambda$-distribution for the stellar component at four different redshifts, divided into spheroidal (red dashed) and disk (blue dot-dashed) galaxies. The stellar component of spheroids has a very low median value, whereas the disk galaxies have a higher one at all shown redshifts.
  }
 \label{fig:lambda_star10+lambda_star}
\end{figure}

The left panels of Figure \ref{fig:lambda_star10+lambda_star} show the $\lambda$-parameter of the stellar component within the innermost 10\% of the virial radius against that of the total DM halo for individual halos classified as spheroids (red circles) and disks (blue diamonds). We can clearly see the split-up of the two populations, where disk galaxies have a higher stellar spin than the spheroids.

On the right side of Figure \ref{fig:lambda_star10+lambda_star} we show the $\lambda$-distribution for the stellar component within the total virial radius. We also see a split-up of the spheroidal (red) and disk (blue) galaxies. The stars in disk galaxies have a higher spin than in spheroids. This seems plausible, since most stars in disk galaxies are found in the disk while most stars in spheroidal galaxies are found in the dominant bulge and thus are more spherically distributed. Hence, they have a net spin that is lower than that in disk galaxies, where the stars sum up their spin. 
This also shows that the stars trace the formation history of the galaxies. 
The stars in disks form out of the fresh gas with high angular momentum, while in spheroids many stars are accreted (via minor/major mergers).

\begin{table}
\caption{K-S Test for the $\lambda_\mathrm{star}$-Distributions}
\label{tab:ks_star}
\centering
\begin{tabular}{l l l}
\hline \hline
  Redshift & $D$ & Probability \\ \hline
  2 & 0.558 & 1.81 $\cdot 10^{-7}$ \\
  1 & 0.364 & 3.42 $\cdot 10^{-5}$ \\
  0.5 & 0.443 & 6.62 $\cdot 10^{-8}$\\
  0.1 & 0.493 & 2.40 $\cdot 10^{-9}$\\
\hline  
\end{tabular}
\tablecomments{The maximum distance $D$ and the significance level (probability) resulting from the K-S test that the $\lambda_\mathrm{star}$-distributions for disks and spheroids at different redshifts are drawn from the same distribution.}
\end{table}

In Table \ref{tab:ks_star} we show the results from the K-S test for the $\lambda_\mathrm{star}$-distribution. The two distributions for the disk and the spheroidal galaxies do not originate from the same distribution. 


In order to see how the $\lambda$-parameter behaves for individual halos in the baryon and the DM-only runs, we identified the same halos in the baryonic and the DM-only run (as described earlier). 
The left panels of the left figure of Figure \ref{fig:lambda_dmo_dmb+jrm} show the $\lambda$-parameter of the DM component against the $\lambda$-parameter of the DM-only run in logarithmic scaling for spheroids at redshifts $z=2$ (upper panel) and $z=0.1$ (lower panel). 
The values of the $\lambda$-parameter of the corresponding halos in both runs are similar. 
The stellar mass within 10\% of the virial radius of the baryon run is color-coded. 
On the right panels we find the $\lambda_\mathrm{DM}$-values for the disks. 
At redshift $z=2$ (upper panel) we note a slight tendency for the $\lambda_\mathrm{DM}$ to decrease with increasing stellar mass. 
The less massive disks tend to have higher $\lambda$-values. 
This is in agreement with \citet{Berta08}, who computed the DM spin parameter of $\approx$52,000 disk galaxies from the SDSS (for details of the sample see references in \citet{Berta08}). 
They found a clear anticorrelation between the DM spin and the stellar mass, i.e., that galaxies with a low mass in general have higher DM spins. 
For the spheroidal galaxies we do not see this trend. 
In addition, this figure demonstrates that a high $\lambda$ does not automatically lead to a disk galaxy and a small $\lambda$ is no guarantee for a spheroidal galaxy. The scatter is equal for both galaxy types. 
The fact that $\lambda_\mathrm{DM}$ in the whole virial radius is similar in the DM-only and the baryon run is in good agreement with \citet{Bryan13}. 


\section{Appendix D: The Radial Specific Angular Momentum Profiles}\label{sec:apprsap}

\begin{figure}
 \begin{centering}
 \centerline{\includegraphics[height=7cm, clip=true]{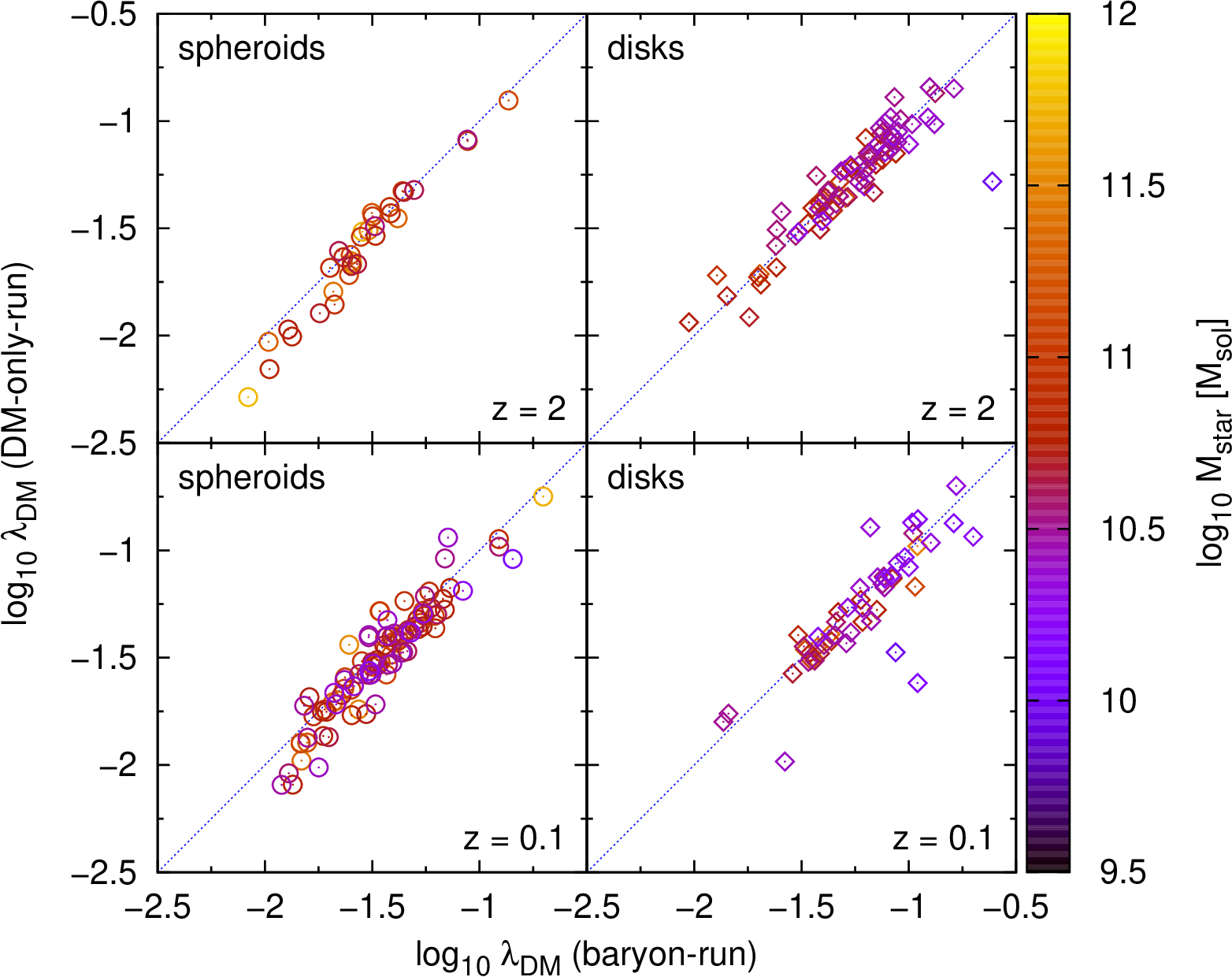}\hspace{1cm}
             \includegraphics[height=7cm, clip=true]{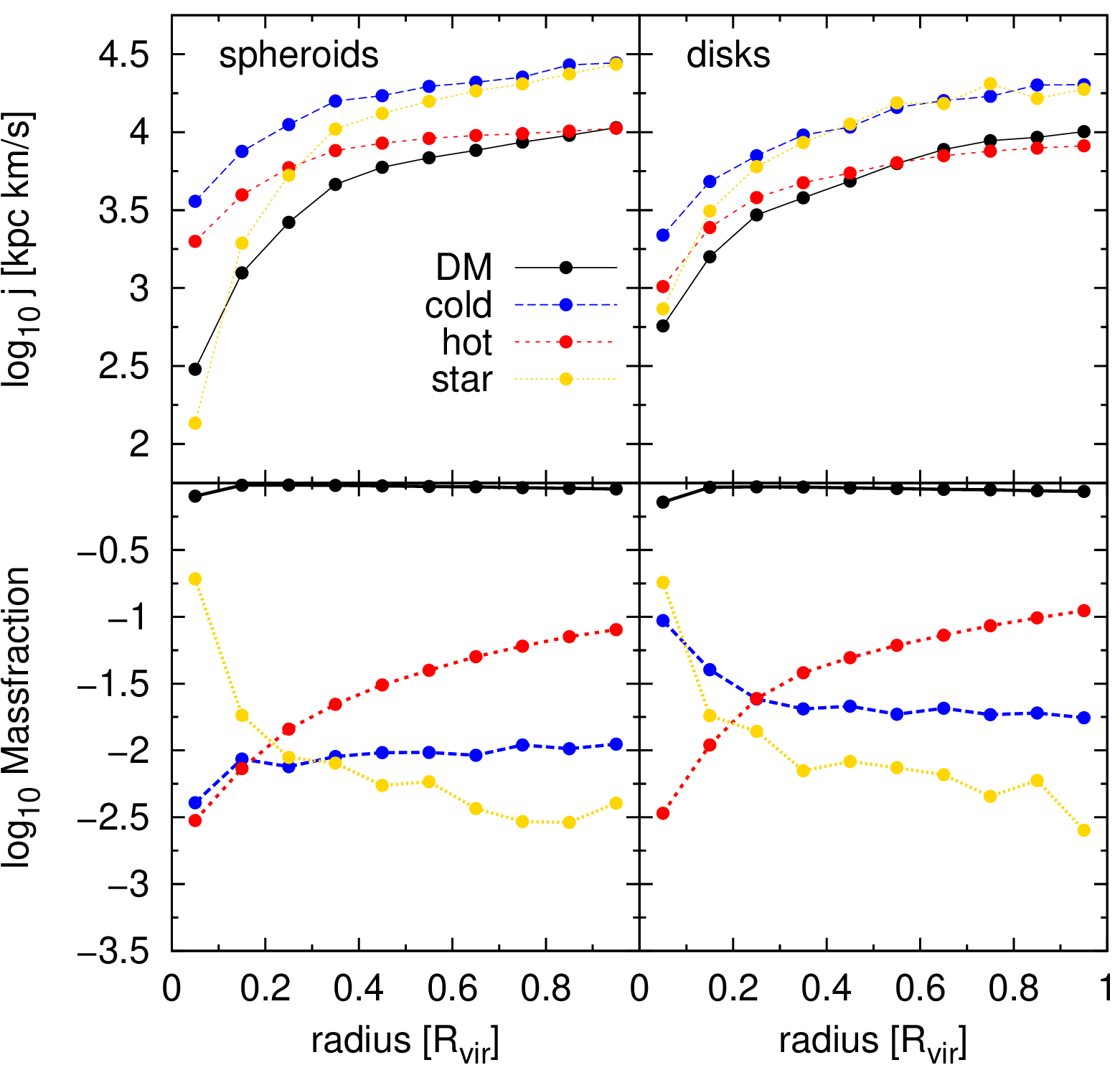}}
 \caption{Left: $\lambda$-parameter plotted logarithmically for the DM component, on the $x$-axis the $\lambda$ for the baryon run and on the $y$-axis for the DM-only run; the color codes the stellar mass within the inner 10\% of the virial radius. On the left panels are the spheroids and on the right panels the disks. The upper panels show redshift $z=2$ and the lower ones $z=0.1$. 
 Right: specific angular momentum (top panels) and the mass fractions (bottom panels) against distance from the center for the cold gas (blue), hot gas (red), stars (yellow), and DM (black) of our sample of galaxies at $z=0.1$. On each left-hand side we show the analysis for the spheroidal galaxies, and on each right-hand side that for the disks.
  }
 \label{fig:lambda_dmo_dmb+jrm}
 \end{centering}
\end{figure}

\begin{figure}
 \begin{centering}
 \centerline{\includegraphics[height=7cm, clip=true]{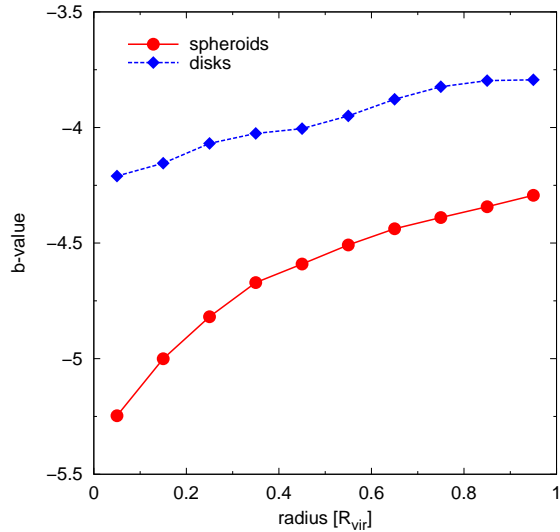}}
 \caption{The radial profile of the $b$-value averaged over spheroids (red circles) and disks (blue diamonds). }
 \label{fig:jrm-bv}
 \end{centering}
\end{figure}

In order to illustrate the behavior of the specific angular momentum of the different components, we show on the right-hand side of Figure \ref{fig:lambda_dmo_dmb+jrm} the mean of our sample of spheroidal (left panel) and disk galaxies (right panel) at $z=0.1$.
In the upper panels we plot the differential specific angular momenta of the different components against radius, and in the lower panels we plot the individual mass fractions of the corresponding components against radius. The $x$-axis is linearly binned with a bin size of 10\% of the virial radius of a halo. 
The specific angular momentum of the two galaxy types increases from the center of the halo up to the virial radius.
On the outer parts the specific angular momentum of the hot gas follows the DM, while the specific angular momentum of the stars follows that of the cold gas. This might be due to the infalling substructures, which mostly consist of small gas clumps with stars that formed out of it. 
The main difference of the two galaxy types becomes clear in the center: the angular momentum of the stars drops dramatically in spheroidal galaxies, while in disk galaxies it remains higher than that of the DM. 
With respect to the baryons, the outer region is dominated by the hot gas, which behaves similarly for spheroids and disks.

Figure \ref{fig:jrm-bv} shows the radial profile of the $b$-value. Therefore, we calculated 
\begin{equation}
 b = \mathrm{log}_{10}\left(\frac{j_\mathrm{star}(\leq r)}{\mathrm{kpc}~\mathrm{km/s}}\right) - \frac{2}{3}\mathrm{log}_{10}\left(\frac{M_\mathrm{star}(\leq r)}{M_{\odot}}\right),
\end{equation}
averaged over the spheroids and disks. The slope is steeper for the spheroids than for the disk galaxies. This illustrates that the measurement of the kinematic properties is more sensitive to the radius for spheroidal than for disk galaxies. 


\bibliography{bibliography,master3,Literaturdatenbank}


\end{document}